\documentclass[twocolumn]{aastex631}   

\usepackage{rotating}
\usepackage{graphicx}
\usepackage{mathptmx}
\usepackage{bm}
\usepackage{esvect}
\usepackage{amsmath}
\usepackage{multirow}
\usepackage{url}
\usepackage{booktabs}
\usepackage[flushleft]{threeparttable}

\submitjournal{ApJ}

\graphicspath{{./}}

\begin{document}

\title{Multiwavelength Analysis of PSR J0437--4715 with Pulse Profile Modeling}

\correspondingauthor{Mingyu GE}
\email{gemy@ihep.ac.cn}
\author{Liqiang QI}
\affiliation{State Key Laboratory of Particle Astrophysics, Institute of High Energy Physics, Chinese Academy of Sciences, Beijing 100049, China}
\author{Juan ZHANG}
\affiliation{State Key Laboratory of Particle Astrophysics, Institute of High Energy Physics, Chinese Academy of Sciences, Beijing 100049, China}
\author{Weiwei Xu}
\affiliation{National Astronomical Observatories (NAOC), Chinese Academy of Sciences, Beijing 100101, China}
\affiliation{School of Physics and Astronomy, Beijing Normal University,  Beijing 100875, China }
\author{Shijie Zheng}
\affiliation{State Key Laboratory of Particle Astrophysics, Institute of High Energy Physics, Chinese Academy of Sciences, Beijing 100049, China}
\affiliation{University of Chinese Academy of Sciences, Chinese Academy of Sciences, Beijing 100049, People’s Republic of China}
\author{Mingyu Ge}
\affiliation{State Key Laboratory of Particle Astrophysics, Institute of High Energy Physics, Chinese Academy of Sciences, Beijing 100049, China}
\affiliation{University of Chinese Academy of Sciences, Chinese Academy of Sciences, Beijing 100049, People’s Republic of China}
\author{Ang Li}
\affiliation{ Department of Astronomy, Xiamen University, Xiamen 361005, People’s Republic of China}
\author{Shuang-Nan Zhang}
\affiliation{State Key Laboratory of Particle Astrophysics, Institute of High Energy Physics, Chinese Academy of Sciences, Beijing 100049, China}
\affiliation{University of Chinese Academy of Sciences, Chinese Academy of Sciences, Beijing 100049, People’s Republic of China}
\author{Hua Feng}
\affiliation{State Key Laboratory of Particle Astrophysics, Institute of High Energy Physics, Chinese Academy of Sciences, Beijing 100049, China}
\affiliation{University of Chinese Academy of Sciences, Chinese Academy of Sciences, Beijing 100049, People’s Republic of China}
\author{Fangjun Lu}
\affiliation{State Key Laboratory of Particle Astrophysics, Institute of High Energy Physics, Chinese Academy of Sciences, Beijing 100049, China}
\affiliation{University of Chinese Academy of Sciences, Chinese Academy of Sciences, Beijing 100049, People’s Republic of China}

\begin{abstract}
We present a multi-wavelength analysis of the nearby millisecond pulsar PSR J0437--4715, combining Hubble Space Telescope (HST) far-ultraviolet, ROSAT soft X-ray, and XMM-Newton X-ray data, to model its broadband emission and energy-resolved pulse profiles, and infer key stellar parameters via Bayesian inference. The broadband emission includes cold thermal, hot thermal, and non-thermal components: cold bulk surface emission is modeled with a non-magnetized partially-ionized hydrogen atmosphere; hot-spot emission adopts the pulse profile modeling technique with a non-magnetized fully-ionized hydrogen atmosphere model; and non-thermal emission is included as a phase-invariant power-law component. By adopting an informative prior on the hot-spot geometry informed by radio polarization position angle measurements, the joint multi-instrument analysis yields a statistically viable and radio-consistent solution with a gravitational mass of 1.38$\pm$0.03~M$_\odot$ and an equatorial circumferential radius of 13.25$_{-0.35}^{+0.34}$~km (68\% confidence intervals). The hot-spot geometry consists of two spherical caps with uniform temperature distributions: the primary hot spot is situated at a colatitude of $\approx$130$^{\circ}$, and the secondary hot spot lies at a colatitude of $\approx$9$^{\circ}$, close to the north pole. It yields tighter radius constraints than HST+ROSAT fits and shifts the radius posterior distribution to larger values relative to NICER-only fits. This work demonstrates the importance of multi-wavelength data in refining neutron star mass-radius measurements and resolving geometric degeneracies. 

\end{abstract}


\section{Introduction}
\label{sec1}
Measurements of neutron star mass and radius provide direct insights into the phase state of dense matter and enable constraints on the Equation Of State (EOS) of matter at supranuclear densities~\citep{lattimer2001neutron}. A leading approach for obtaining precise constraints on the compactness $M/R$ involves modeling pulsed thermal emission from heated surface regions of rotation-powered millisecond pulsars---known as the pulse profile modeling (PPM) technique~\citep{pavlov1997mass,Zavlin1998,poutanen2003nature,poutanen2006pulse,cadeau2007light,morsink2007oblate,algendy2014universality,nattila2018radiation,bogdanov2019constrainingii}. Given well-constrained masses, orbital inclinations, and distances from radio timing observations, this technique yields tightly constrained values for the neutron star radius. Recent studies have applied the PPM technique to data from the Neutron Star Interior Composition Explorer (NICER)~\citep{gendreau2016neutron} for a suite of nearby rotation-powered millisecond pulsars, including PSR J0030+0451~\citep{miller2019psr,riley2019nicer,salmi2023atmospheric,vinciguerra2024updated}, PSR J0740+6620~\citep{miller2021radius,riley2021nicer,salmi2022radius, salmi2024radius,dittmann2024more}, PSR J0437--4715~\citep{choudhury2024nicer}, PSR J1231--1411~\citep{salmi2024nicer,Qi_2025}, and PSR J0614--3329~\citep{mauviard2025nicer}. An expanded library of mass-radius ($M$-$R$) measurements serves as a critical benchmark for constraining the dense-matter EOS~\citep[e.g.][]{watts2016colloquium,miller2019psr,raaijmakers2021constraints,miao2024thermal}. Additionally, inferred hot-spot geometries offer clues to a neutron star's magnetic field structure and X-ray emission mechanism~\citep{bilous2019nicer,chen2020numerical,kalapotharakos2021multipolar,carrasco2023relativistic}.

This work focuses on PSR J0437--4715, the nearest and brightest millisecond pulsar. Its proximity and brightness facilitate precise measurements of key parameters (mass, orbital inclination, and distance) via radio pulsar timing~\citep{Reardon_2024}, as well as comprehensive multi-wavelength observational data spanning the infrared to gamma-ray bands~\citep{Abdo_2010,Durant_2012,Bogdanov_2013}. Regarding radius measurements for this pulsar, modeling of XMM-Newton pulse profiles (0.5-1.8~keV) yields a lower limit on the neutron star radius of $>$11.1~km at 3$\sigma$ confidence level, assuming a neutron star mass of 1.76~M$_\odot$~\citep{Bogdanov_2013}. Recent radio timing measurements---with a mass estimate of 1.418$\pm$0.044~M$_\odot$~\citep{Reardon_2024}---slightly shift this constraint toward smaller radii. Additionally, a joint fit of phase-averaged Hubble Space Telescope (HST) and ROSAT data (far ultraviolet, FUV; soft X-ray, 0.1-0.4~keV) yields a radius of 13.1$_{-0.7}^{+0.9}$~km, assuming that the radiation originates from the thermal emission of the entire cold stellar surface~\citep{10.1093/mnras/stz2941}. The posteriors shift to 13.6$_{-0.8}^{+0.9}$~km when incorporating contributions from heated regions (modeled as blackbody spectra). Most recently, \cite{choudhury2024nicer} apply the PPM technique to NICER data (0.3-3.0 keV) and infer a gravitational mass of $M$~=~1.418$\pm$0.037~M$_\odot$ and an equatorial circumferential radius of $R$~=~11.36$_{-0.63}^{+0.95}$~km. Their analysis models hot-spot thermal emission using a background-marginalized likelihood function (to account for phase-invariant components unrelated to the hot spots) and explores various parameterized geometries. Their headline result is the CST+PDT model, i.e.\ a Concentric Single Temperature spot and a Protruding Dual Temperature spot~\citep{choudhury2024nicer}.

Multi-wavelength observations and studies indicate that PSR J0437--4715's emission comprises multiple components: thermal emission from both the entire stellar surface and heated polar caps, and non-thermal emission---likely from the pulsar's magnetosphere and/or a faint wind nebula \citep{Durant_2012,Bogdanov_2013,10.1093/mnras/stz2941,10.1093/mnras/stw2194}. In this work, a joint fit is performed to the energy spectra from HST and ROSAT data and energy-resolved pulse profiles from XMM-Newton data using the PPM technique, incorporating both thermal and non-thermal emission components. This analysis aims to reproduce the broadband emission characteristics, constrain the neutron star radius, and infer plausible hot-spot geometries. The structure of the paper is organized as follows: the data processing of XMM-Newton, ROSAT, HST, and NICER observations is presented in Section~\ref{sec2}; the methodology is validated in Section~\ref{sec3}, including cold bulk surface emission modeling and hot-spot pulse profile modeling; the inferred model parameters of the multi-instrument fit are presented and discussed in Section~\ref{sec4}; and the conclusions are given in Section~\ref{sec5}.

\section{Observation data processing}
\label{sec2}

\subsection{XMM-Newton}
The observation data of XMM-Newton~\citep{jansen2001xmm} EPIC-pn~\citep{struder2001european} are used in this work (ObsID 0603460101). The timing mode of EPIC-pn with a time resolution of 30~$\mu$s enables the extraction of energy-resolved pulse profiles---critical for investigating thermal emission from heated regions of PSR J0437--4715. EPIC-mos data are excluded because they yield only phase-averaged energy spectra (insufficient for pulse profile modeling). Data processing is performed using the XMM-Newton Science Analysis System (SAS, version 18.0.0) and the calibration database (Update: 2023 May 10). Observation data files are converted to event files via the tool \textit{epproc}. To mitigate the impact of background flares, an additional time cut (TIME in [3.77301e8:3.77405e8]) is applied alongside the standard SAS selection criteria (PATTERN$\leq$4) \&\& (PI in [250:15000]) \&\& \#XMMEA\_EP \&\& (FLAG==0); the filtering reduces the effective exposure to 101436~s. Source events are extracted from the region defined by (RAWX in [29:43]) \&\& (RAWY$>$13), while background events are extracted from the region defined by !(RAWX in [29:43]) \&\& (RAWY$>$13) \&\& (RAWX!=45) \&\& (RAWX!=61) \&\& (RAWX!=27) \&\& (RAWX!=19) \&\& (RAWX!=21) \&\& (RAWX!=25). The scaling factor for the background spectra is calculated to be 0.35, whereas direct application of the SAS tool \textit{backscal} results in an underestimation of this value. The SAS tools \textit{arfgen} and \textit{rmfgen} are used to calculate the instrument response files---specifically, Ancillary Response Files (ARFs) and Redistribution Matrix Files (RMFs). The net count rate in the 0.3-3.0~keV energy range is $\approx$0.55~cts/s. A lower threshold of 0.3~keV is adopted to maximize the statistics, and an upper limit of 3.0~keV is set, as the spectrum is dominated by particle background above this energy. Calibration uncertainties in the EPIC-pn timing mode below 0.5~keV are discussed in Section~\ref{sec3.3}.

\begin{figure}[h]
\begin{center}
\includegraphics[width=0.9\linewidth]{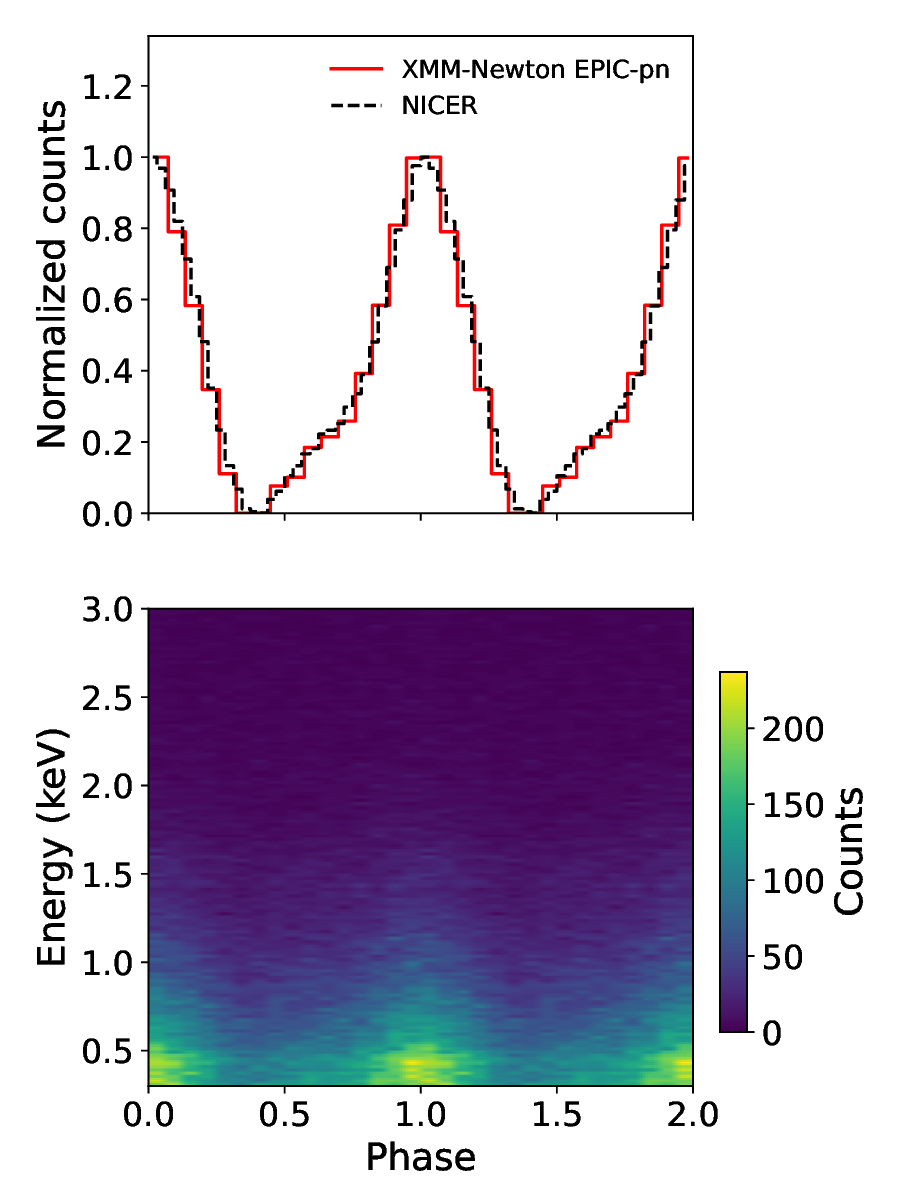}
\end{center}
\caption{Normalized energy-integrated pulse profiles of PSR J0437--4715 in the 0.3-3.0 keV band, extracted from XMM-Newton EPIC-pn (16 phase bins) and NICER data (32 phase bins; upper panel). Energy-resolved pulse profiles from XMM-Newton EPIC-pn data, spanning 0.3-3.0 keV with 16 phase bins and 135 energy channels (lower panel). Two rotational cycles are plotted for clarity.}
\label{obsdata}
\end{figure}

To determine the pulsar rotation phase for each filtered event, event times are converted to the solar system barycenter~\citep{2024Univ...10..174Z} based on the ephemeris of PSR J0437--4715 from the International Pulsar Timing Array Data Release 2 (IPTA DR2)~\citep{2019MNRAS.490.4666P} and then folded with the timing parameters of this ephemeris. The normalized energy-integrated pulse profile (0.3-3.0~keV, 16 phase bins) is shown in the upper panel of Figure~\ref{obsdata}, alongside the energy-integrated pulse profile extracted from NICER data (0.3-3.0~keV, 32 phase bins)~\citep{J0437data}; good agreement between the two is evident. The energy-integrated pulse profile consists of a broad, prominent main pulse peak and a diffuse hump near rotation phase 0.55. Owing to limited statistics in the energy-resolved pulse profiles, the energy spectrum is re-binned into groups of 4 channels (see the lower panel of Figure~\ref{obsdata}) and the RMF is re-binned correspondingly. The tailored response files, energy-resolved pulse profiles, and scaled background spectrum are then used as inputs for subsequent PPM. 

\subsection{ROSAT}
The ROSAT PSPC observations~\citep{TRUMPER1982241} are used to complement the XMM-Newton EPIC-pn data of PSR J0437--4715 (ObsID 701184). These data provide reliable spectral information down to 0.1~keV with well-characterized calibration. Event extraction from archival data is performed via the tool \textit{xselect}. Following the procedure of \cite{10.1093/mnras/stw2194}, source events are selected within a 70-arcsecond radius circle centered at R.A.(J2000)~=~69.3188829$^{\circ}$, Dec.(J2000)~=~-47.2508931$^{\circ}$; background events are extracted from a source-centered annulus with inner and outer radii of 70 and 110 arcseconds, respectively. The ARF is generated using the tool \textit{pcarf}. The on-axis RMF is obtained from the Rutgers University FTP server\footnote{\url{https://www.physics.rutgers.edu/~matilsky/documents/pspcb_gain2_256.rmf}}. The net count rate in the 0.1-0.5~keV energy range is $\approx$0.13~cts/s with an effective exposure of 5957~s. 

\subsection{HST}
To constrain thermal emission from the cold bulk surface of PSR J0437--4715, HST FUV flux measurements are used. Specifically, the FUV fluxes and their associated uncertainties with wavelength bin centers spanning 1245-1695~$\AA$ are adopted (Table~3 in \cite{Durant_2012}). The last three FUV flux bins are excluded from the analysis because they exhibit a narrow peak structure of unknown origin~\citep{Durant_2012}. Near ultraviolet fluxes are excluded, as they are dominated by emission from the pulsar's white-dwarf companion. 

\subsection{NICER}
The NICER XTI~\citep{gendreau2016neutron} observations of PSR J0437--4715 are used to validate the independently developed PPM numerical algorithm presented in this work. The 3C50 dataset and corresponding response files from \cite{J0437data} are adopted, which include the energy-resolved pulse profiles, background estimates, mean spectrum of the nearby Active Galactic Nucleus (AGN) RX J0437.4--4711, ARF, and RMF. For this dataset, the net count rate in the 0.3-3.0~keV energy range is $\approx$0.95~cts/s with an effective exposure of 1.328~Ms. 

\section{Methodology}
\label{sec3}

\subsection{HST+ROSAT}
\label{sec3.1}
A Rayleigh-Jeans tail is detected in the HST FUV data of PSR J0437--4715 in excess of the white-dwarf companion emission~\citep{Durant_2012}. Notably, the XMM-Newton energy spectrum cannot be adequately described by spectral models of BB($\times$2)+PL, Hatm($\times$2)+PL, or CBB($\times$2)~\citep{Bogdanov_2013}. These observational features imply the presence of thermal emission from the neutron star's entire cold surface. \cite{10.1093/mnras/stz2941} perform a simultaneous fit to the FUV and soft X-ray spectra using a non-magnetized partially-ionized atmospheric model, which favors a hydrogen composition. In this work, an independent numerical algorithm is implemented to calculate the emergent spectrum of the neutron star's cold atmosphere; the resulting synthetic spectra are tabulated for subsequent Bayesian analysis to constrain the neutron star radius. To validate both the atmospheric modeling and the Bayesian parameter estimation framework, this section revisits the joint fit of the HST FUV and ROSAT soft X-ray data. 

\begin{figure}[h]
\begin{center}
\includegraphics[width=0.8\linewidth]{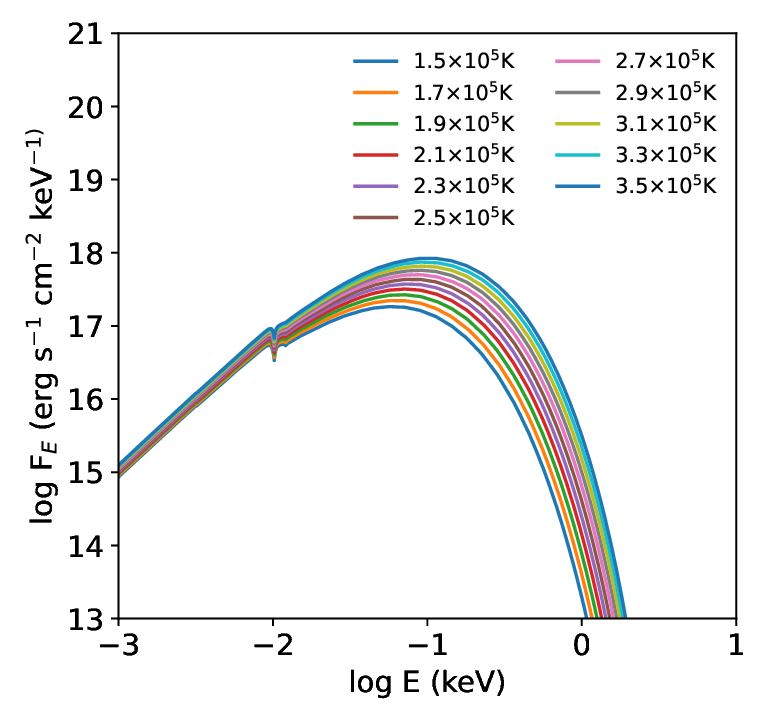}
\end{center}
\caption{Non-redshifted energy spectra of a non-magnetized partially-ionized hydrogen atmosphere, computed for an effective surface gravity of $g_\textup{eff}$~=~2~$\times$10$^{14}$~cm~s$^{-2}$ and effective bulk temperature $T_\textup{eff,bulk}$ ranging from 1.5$\times$10$^5$~K to 3.5$\times$10$^5$~K. }
\label{cold_spectra}
\end{figure}

Following the prescription of \cite{10.1093/mnras/stz2941}, a non-magnetized partially-ionized hydrogen atmosphere is adopted for the cold emission from PSR J0437--4715. The effective bulk surface temperature of this pulsar is estimated to be in the range (1.25-3.5)$\times$10$^5$~K~\citep{Durant_2012}, a regime where scattering processes can be neglected in the modeling. For a neutron star atmosphere of centimeter-scale thickness, the problem simplifies to a 1D plane-parallel radiative transfer calculation. Under the assumption of Local Thermodynamic Equilibrium (LTE), the formal radiative transfer equation can be solved with a second-order integration in the upward direction as follows~\citep{OLSON1987325}, 
\begin{equation}
I_{i+1/2} = e^{-\Delta \tau_i} I_{i-1/2} + Q_i \\,
\end{equation}
with
\begin{equation}
Q_i = \Big(\frac{1-(1+\Delta \tau_i)e^{-\Delta \tau_i}}{\Delta \tau_i}\Big) S_{i-1/2} + \big(\frac{\Delta \tau_i -1 +e^{-\Delta \tau_i}}{\Delta \tau_i}\big)  S_{i+1/2} \\,
\end{equation}
where $I_{i-1/2}$ and $I_{i+1/2}$ denote the specific intensity at the lower and upper boundaries of the cell $i$; $\Delta \tau_i$ is the optical depth of cell $i$; and $S_{i-1/2}$ and $S_{i+1/2}$ represent the source function at the cell boundaries (equal to the Planck function in LTE). The boundary conditions of the problem are set with the deepest layer in a thermal equilibrium with the matter and no incoming radiation at the surface. The downward integration is performed analogously by changing the indices in the formula. Given an initial temperature profile of the atmosphere, the radiative transfer equation is solved iteratively: the Lucy-Uns\"old correction is applied to refine the temperature profile until the energy balance is achieved~\citep{1978stat.book.....M}, i.e.\ the bolometric flux emergent from the atmosphere matches the target flux of $\sigma T_\textup{eff,bulk}^4$ ($\sigma$ being the Stefan-Boltzmann constant, and $T_\textup{eff,bulk}$ being the effective bulk surface temperature). The non-redshifted energy spectra for a grid of effective surface gravities and temperatures are computed and tabulated, with the energy-dependent opacities retrieved from the Los Alamos Opacity Project\footnote{\url{https://aphysics2.lanl.gov}}. An illustrative example of these non-redshifted spectra is presented in Figure~\ref{cold_spectra}. 

The joint spectral fit is conditional on informative tight priors for the neutron star mass and distance from high-precision radio pulsar timing measurements~\citep{Reardon_2024}; the neutron star mass is assigned a Gaussian prior Probability Density Function (PDF), and the distance is fixed $\approx$156.980~pc. The neutron star radius is treated as an uninformative parameter, with a uniform prior PDF spanning 9-17~km. The effective bulk surface temperature is assigned a uniform prior PDF over (1.5-3.5)~$\times$10$^5$~K. To account for the propagation effects when a photon travels through the interstellar medium of galaxies, the Milky Way extinction curve of \cite{Clayton_2003} with a polynomial function from \cite{1990ApJS...72..163F} is used for the HST FUV data (fixing the extinction ratio $R_V$ to 3.1). The interstellar dust extinction $E(B-V)$ is assigned a Gaussian prior PDF~\citep{gaia}, truncated to the physically motivated range 0.00001$<E(B-V)<$0.07. Given the neutron star mass, radius, and distance, the observed FUV data are de-reddened and directly compared to the synthetic atmospheric models. A Gaussian likelihood function is employed to quantify the probability of observing the data given the model parameters. 

For the X-ray band, interstellar absorption is modeled using the TBabs model~\citep{wilms2000absorption}; the energy-dependent photo-absorption cross-sections are extracted from the XSPEC software package~\citep{1999ascl.soft10005A} and tabulated for subsequent interpolation. The neutral hydrogen column density $N_\textup{H}$ is assigned a uniform prior PDF over the range (0.004-2)~$\times$10$^{20}$~cm$^{-2}$. To enable direct comparison with the observed X-ray count rates, the emergent spectrum incident on the telescope aperture is convolved with the instrument response files (ARFs and RMFs). An energy-independent scaling coefficient $\alpha$ is introduced for the effective area curve (assigned a Gaussian prior PDF) to account for residual calibration uncertainties. Owing to the low count statistics in individual ROSAT spectral channels, a Poisson likelihood function with a known background is adopted for X-ray data fitting. A complete summary of all prior PDFs and their specifications is provided in Table~\ref{J0437coldtable}. 

\begin{figure}[h]
\begin{center}
\includegraphics[width=0.85\linewidth]{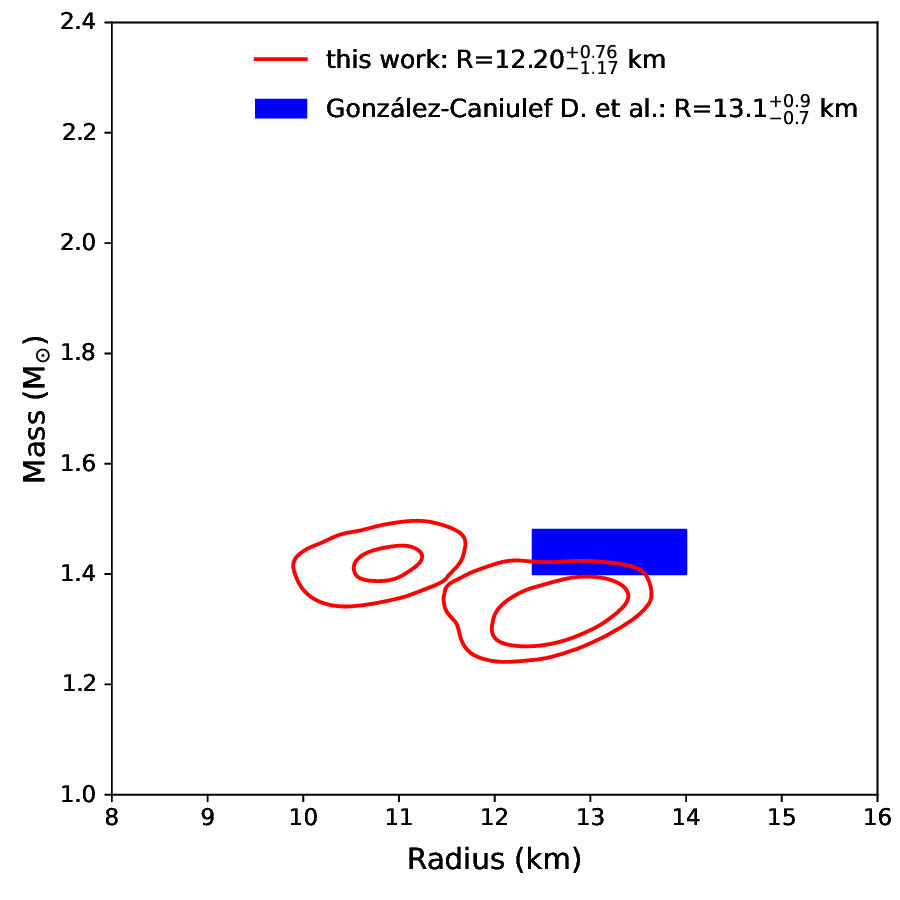}
\end{center}
\caption{Two-dimensional marginalized posterior PDFs of neutron star mass and radius, derived from the joint HST+ROSAT fit excluding hot-region contributions. The contours represent the 68\% and 95\% credible regions. Results are compared to the measurements of \cite{10.1093/mnras/stz2941}. The shaded area represent the 68\% credible region. }
\label{J0437coldfigure}
\end{figure}

With the prior PDFs and likelihood functions fully defined, the publicly available nested sampler MultiNest~\citep{feroz2009multinest} is employed to perform the Bayesian parameter estimation and model evaluation. The default MultiNest settings adopted for most runs in this work are summarized in Table~\ref{MultiNest}. If multi-modal structures are identified in the posterior parameter space during post-processing, the multi-mode option is enabled, and the Bayesian analysis is re-run with an increased number of live points to ensure thorough exploration of the parameter space. The inferred parameters with 68\% confidence intervals (CIs) and best-fit values are listed in Table~\ref{J0437coldtable}. The two-dimensional marginalized posterior PDFs of neutron star masses and radii are shown in Figure~\ref{J0437coldfigure}. The inferred radius posteriors exhibit an adjacent yet disjoint distribution at lower radii, which cannot be resolved by MultiNest’s default separation algorithm. They are in good agreement with the result of \cite{10.1093/mnras/stz2941} (R~=~13.1$_{-0.7}^{+0.9}$~km), demonstrating a proper implementation of the hydrogen atmosphere modeling and Bayesian parameter estimation. In this section, the contributions from the heated regions and non-thermal emission are excluded from the fit; these components are incorporated and discussed in detail in Section~\ref{sec4}. 

\begin{table*}[ht]
\centering
\begin{threeparttable}
\caption{Summary of priors and Bayesian parameter estimates from the joint fit of HST FUV and ROSAT soft X-ray data, adopting a non-magnetized partially-ionized hydrogen atmosphere model.}
\begin{tabular}{lccccc}
\hline
Parameter		& Description    &Prior &$\widehat{\textup{CI}}_{68\%}$ & Best-fit \\
\hline
$M$ (\(\textup{M}_\odot\))& gravitational mass 					
&$M\sim$~Gaussian(1.418,0.044)			& 1.35$_{-0.05}^{+0.06}$	&1.32\\
$R_\textup{eq}$ (km)& equatorial circumferential radius 						
&$R_\textup{eq}\sim$~Uniform(9, 17)		& 12.20$_{-1.17}^{+0.76}$	&12.85\\
$D$ (pc)& distance to Earth 						
&$D\sim$~Uniform(156.979, 156.981)		& 156.980$\pm$0.001		&156.981\\
$N_\textup{H}$ (10$^{20}$ cm$^{-2}$)& neutral hydrogen column density			
&$N_\textup{H}\sim$~Uniform(0.004, 2)		& 0.79$_{-0.18}^{+0.17}$			&0.84\\
$E(B-V)$ & interstellar dust extinction 					
&$E(B-V)\sim$~Gaussian(0.005,0.003) 		& 0.007$\pm$0.003		&0.014\\
&&{truncated between 0.00001 and 0.07}& && \\
\hline
$\alpha_\textup{ROSAT}$& ROSAT effective area scaling 				
&$\alpha_\textup{ROSAT}\sim$~Gaussian(1, 0.05)	& 1.02$\pm$0.05			&1.06\\
\hline
$T_\textup{eff,bulk}$(10$^{5}$ K)& effective bulk surface temperature 		
&$T_\textup{eff,bulk}\sim$~Uniform(1.5, 3.5)	& 2.50$_{-0.16}^{+0.18}$		&2.44\\
\hline
\end{tabular}
\label{J0437coldtable}
\end{threeparttable}
\end{table*}

\begin{table}[h]
\centering
\caption{Default configuration settings for the MultiNest nested sampling algorithm used in Bayesian inference runs of this work.}
\begin{tabular}{lc}
\hline
Parameter & Value \\
\hline
Importance nested sampling & off \\
Evidence tolerance & 0.1 \\
Sampling efficiency  & 0.01 \\
Multi-mode & off \\
Live points & 1000\\
\hline
\end{tabular}
\label{MultiNest}
\end{table}

\subsection{NICER-only}
\label{sec3.2}
In this section, the pulse profile modeling and Bayesian parameter estimation applied to the NICER-only data are briefly presented. An independent C++ numerical algorithm~\citep{Qi_2025} based on the Schwarzschild spacetime and the Doppler approximation is employed in this work to compute the thermal emission from heated regions on the surface of PSR J0437--4715. While previous work models emission using the spherical-star Schwarzschild spacetime and the Doppler approximation~\citep{Qi_2025}, the framework is extended to account for the oblateness effect of fast-rotating neutron stars, following the prescription of \cite{morsink2007oblate}. For PSR J0437--4715 (spin frequency 173.69~Hz), the oblateness effect is still taken into account, notwithstanding its negligible magnitude. A validation of this numerical algorithm is performed by comparing its results with those of \cite{choudhury2024nicer}, which are derived using the open-source X-ray Pulse Simulation and Inference software package (X-PSI)~\citep{riley2023x}.

As discussed in Section~\ref{sec3.1}, the surface of PSR J0437--4715 is highly likely to be covered by a hydrogen atmosphere. For hot regions (with effective temperatures of order 10$^6$~K, significantly higher than the cold bulk surface), a non-magnetized fully-ionized hydrogen atmosphere model is adopted (utilizing the lookup table of nsx\_H\_v200804.out)~\citep{nsxmodel}---in contrast to the partially-ionized model used for the cold bulk surface emission. Though other light-element atmospheric compositions cannot be entirely ruled out, the hydrogen atmosphere model has been successfully applied to multiple NICER targets~\citep{miller2019psr,riley2019nicer,miller2021radius,riley2021nicer,choudhury2024nicer,salmi2024nicer,Qi_2025,mauviard2025nicer}, reproducing the observed energy-resolved pulse profile. Photon propagation from the stellar surface to the telescope includes relativistic effects due to the strong gravitational field and rapid rotation of neutron stars, attenuation by the interstellar medium, and convolution with the instrumental response. Relativistic ray-tracing is performed in the Oblate-star Schwarzschild spacetime and the Doppler approximation following the detailed formalism of \cite{poutanen2003nature,poutanen2006pulse,cadeau2007light,morsink2007oblate,algendy2014universality,nattila2018radiation,bogdanov2019constrainingii}. The interstellar attenuation model and the instrumental convolution are identical to those described in Section~\ref{sec3.1}. 

Regarding hot-spot geometry, \cite{choudhury2024nicer} conduct an exhaustive analysis of phenomenological models, including circular regions, annular rings, crescents, and combinations thereof. Starting from the simplest model of two single-temperature circular regions (i.e.\ Single Temperature -- Unshared, ST-U), the model consisting of one single-temperature annulus and one dual-temperature overlapping circular region (CST+PDT) is proposed as the headline result. This CST+PDT model adequately describes the NICER data and is insensitive to background constraints~\citep{choudhury2024nicer}. Physically, this hot-spot geometry may be explained by an offset dipole magnetic field or the presence of quadrupole field components~\citep{10.1093/mnras/stz2524}.

In Bayesian inference of NICER-only data, bulk surface emission and non-thermal emission are not explicitly accounted for in the model. Instead, a background-marginalized likelihood function is employed to capture phase-invariant components external to the hot spots, i.e.\ to integrate the likelihood over possible background count rates in each energy channel. Phase-invariant components include thermal emission from the neutron star's cold bulk surface, non-thermal emission, instrumental background, and flux contamination from sources within the NICER field of view. For PSR J0437--4715, contamination from the nearby bright AGN RX J0437.4--4711 has a substantial impact on the inferred results. Additionally, time-dependent background estimation for NICER is subject to significant uncertainties due to its non-focusing optics, non-imaging detector, and complex radiation environments. The upper and lower limits on the background constraints are carefully treated, accounting for both the AGN spectrum and instrumental background estimation~\citep{choudhury2024nicer}. 

\begin{figure}[h]
\begin{center}
\includegraphics[width=0.85\linewidth]{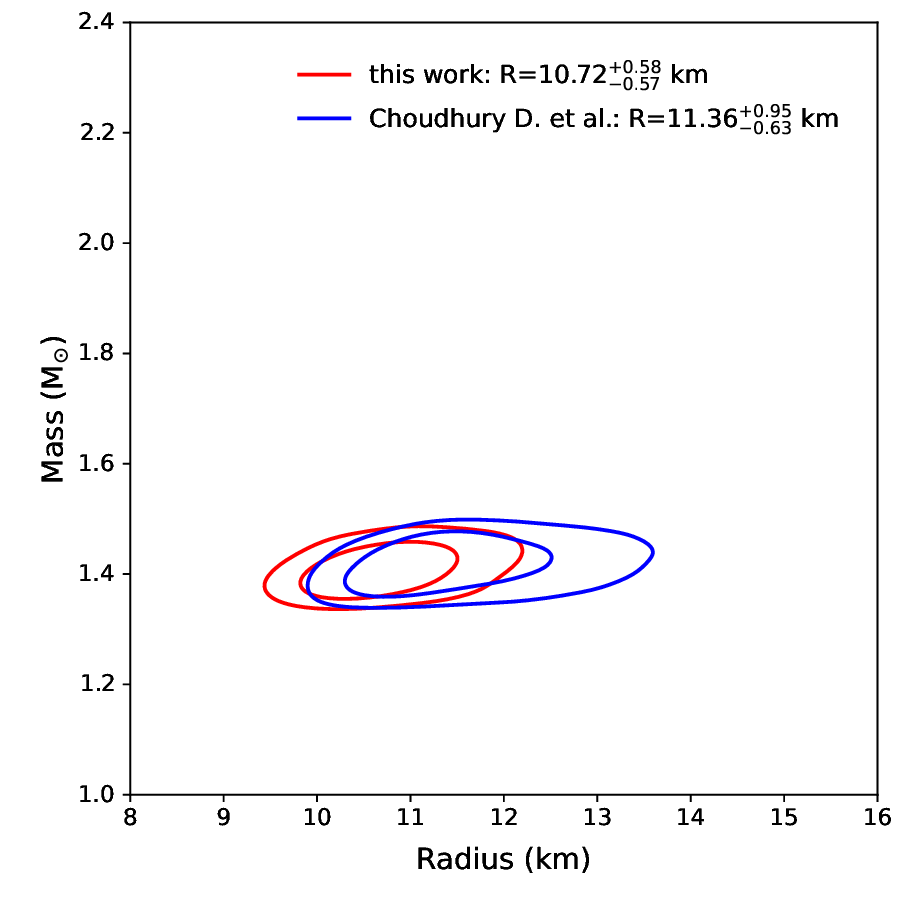}
\end{center}
\caption{Two-dimensional marginalized posterior PDFs of neutron star mass and radius from the NICER-only fit. Results are compared to the headline measurements of \cite{choudhury2024nicer}. The contours represent the 68\% and 95\% credible regions. }
\label{niceronlyfig}
\end{figure}

To validate the independently developed PPM numerical algorithm, the Bayesian inference is performed using the same dataset and response files from \cite{J0437data}, adopting the CST+PDT hot-spot geometry. The lower and upper background bounds are defined as $\max(0,\{B_\textup{3C50}\}-3\times\{\sigma_\textup{3C50}\}+0.25\times\{B_\textup{AGN}\})$ and $(\{B_\textup{3C50}\}+3\times\{\sigma_\textup{3C50}\}+2\times\{B_\textup{AGN}\})$, respectively, where $\{ B_\textup{3C50}\}$ and $\{\sigma_\textup{3C50}\}$ denote the instrumental background estimate and its uncertainty for the 3C50 dataset, and $\{B_\textup{AGN}\}$ is the mean flux from the contaminating AGN. The MultiNest settings are identical to those outlined in Section~\ref{sec3.1}; the adopted prior PDFs and resulting Bayesian parameter estimates are summarized in Table~\ref{niceronlytable}. The inferred neutron star radius of 10.72$_{-0.57}^{+0.58}$~km is consistent with the result of \cite{choudhury2024nicer} ($R$~=~11.36$_{-0.63}^{+0.95}$~km) at the CI$_{68}$ level (see Figure~\ref{niceronlyfig}), demonstrating a proper implementation of the numerical algorithm in this work. The discrepancies between the two results likely arise from differences in PPM resolution settings, MultiNest configuration, and other algorithmic specifications (e.g.\ hot-spot mesh discretizations). 

\begin{table*}[ht]
\centering
\begin{threeparttable}
\caption{Summary of priors and Bayesian parameter estimates from the fit of NICER-only energy-resolved pulse profiles, adopting a non-magnetized fully-ionized hydrogen atmosphere model for hot-region emission.}
\begin{tabular}{lccccc}
\hline
Parameter	& Description 	&Prior &$\widehat{\textup{CI}}_{68\%}$ & Best-fit \\
\hline
$F_0$ (Hz)& spin frequency		 							
&173.69, fixed					& -	& - \\

$M$ (\(\textup{M}_\odot\))& gravitational mass 					
&$M\sim$~Gaussian(1.418,0.044)			& 1.41$_{-0.04}^{+0.03}$			&1.45\\

$R_\textup{eq}$ (km)& equatorial circumferential radius 						
&$R_\textup{eq}\sim$~Uniform(9, 17)		& 10.72$_{-0.57}^{+0.58}$		&11.71\\

$\cos(i)$& cosine of view inclination 					
&$\cos(i)\sim$~Gaussian(-0.7373,0.0002)		& -0.7373$\pm$0.0002		&-0.7374\\

$D$ (pc)& distance to Earth 						
&$D\sim$~Uniform(156.979, 156.981)		& 156.980$\pm$0.001		&156.979\\

$N_\textup{H}$ (10$^{20}$ cm$^{-2}$)& neutral hydrogen column density			
&$N_\textup{H}\sim$~Uniform(0.004, 2)		& 0.06$_{-0.04}^{+0.05}$			&0.05\\

\hline
$\alpha_\textup{NICER}$& NICER effective area scaling 				
&$\alpha_\textup{NICER}\sim$~Gaussian(1, 0.05)	& 1.00$\pm$0.04			&1.04\\

\hline
$\theta_\textup{p}$ (deg)& center colatitude of primary spot 			
&$\cos(\theta_\textup{p})\sim$~Uniform(-1, 1)	& 121.73$_{-2.55}^{+2.58}$		&124.00\\

$\Delta\theta_\textup{p}$ (deg)& angular radius of primary spot 		
&$\Delta\theta_\textup{p}\sim$~Uniform(0.01, 45)& 10.67$_{-0.81}^{+0.76}$		&10.42\\

$kT_\textup{eff,p,ceding}$ (keV)& effective temperature of primary spot			
&$kT_\textup{eff,p,ceding}\sim$~Uniform(0.011, 0.3)& 0.053$\pm$0.003		&0.049\\

$kT_\textup{eff,p,superceding}$ (keV)& effective temperature of superceding 	
&$kT_\textup{eff,p,superceding}\sim$~Uniform(0.011, 0.3)& 0.149$\pm$0.004	&0.145\\

$f_\textup{p}$	& angular radius ratio of primary spot		
&$f_\textup{p}$ $\sim$~Uniform(0.001, 1.999)	& 1.60$_{-0.27}^{+0.24}$			&1.85\\

$\varkappa_\textup{p}$& angular separation of primary spot  		
&$\varkappa_\textup{p}$ $\sim$~Uniform(0, 1)	& 0.59$_{-0.41}^{+0.30}$			&0.17\\

$\varphi_\textup{p}$ (deg) 	& azimuthal offset of primary spot		
&$\varphi_\textup{p}$ $\sim$~Uniform(0, 360)	& 181.20$_{-2.19}^{+2.29}$		&176.79\\

$\phi_\textup{p}$	& center phase of primary spot			
&$\phi_\textup{p}\sim$~Uniform(a+0.4, a+0.7)$^*$	& a+0.519$\pm$0.005		&a+0.517\\

$\theta_\textup{s}$ (deg)& center colatitude of secondary spot 			
&$\cos(\theta_\textup{s})\sim$~Uniform(-1, 1)	& 8.08$_{-0.68}^{+0.67}$			&7.04\\

$\Delta\theta_\textup{s}$ (deg)	& angular radius of secondary spot 		
&$\Delta\theta_\textup{s}\sim$~Uniform(0.01, 45)& 23.20$_{-2.51}^{+2.53}$		&25.52\\

$kT_\textup{eff,s}$ (keV)	& effective temperature of secondary spot 		
&$kT_\textup{eff,s}\sim$~Uniform(0.011, 0.3)	& 0.112$\pm$0.003		&0.109\\

$f_\textup{s}$	& angular radius ratio of secondary spot		
&$f_\textup{s}$ $\sim$~Uniform(0, 1)		& 0.17$\pm$0.11			&0.19\\

$\phi_\textup{s}$	& center phase of secondary spot 			
&$\phi_\textup{s}\sim$~Uniform(a-0.1, a+0.3)	& a+0.050$\pm$0.005		&a+0.050\\

\hline
\end{tabular}
\begin{tablenotes}
\small \item $^*$The variable of a denotes a source-specific phase offset, implemented to facilitate pulse profile modeling.
\end{tablenotes}
\label{niceronlytable}
\end{threeparttable}
\end{table*}

\section{Results and Discussion}
\label{sec4}

\subsection{HST+ROSAT+XMM}
\label{sec3.3}

The radiation emitted by PSR J0437--4715 comprises multiple components: thermal emission from the cold bulk surface and heated region, and non-thermal emission likely from the magnetosphere and/or faint wind nebula. In this section, a joint fit of the multi-wavelength spectra and energy-resolved pulse profiles is performed using observations from HST, ROSAT, and XMM-Newton. For the FUV and soft X-ray spectral fitting, the non-magnetized partially-ionized hydrogen atmosphere model is adopted. For the X-ray energy-resolved pulse profile fitting, the PPM technique with a non-magnetized fully-ionized hydrogen atmosphere model is used to characterize the hot-region thermal emission.

Hot-spot contributions are excluded from the FUV band analysis; the validity of this assumption is discussed in Section~\ref{sec4.3}. On the other hand, the low-energy tail of the hot-region emission can extend down to 0.1~keV in the X-ray band, so its contribution is included when fitting the ROSAT data. In contrast to the blackbody spectra used by \cite{10.1093/mnras/stz2941}, the synthetic spectra from the PPM framework with a hydrogen atmosphere are used. Similarly, the high-energy tail of the cold bulk surface emission can extend up to 1.0~keV in the X-ray band, depending on its effective temperature (see Figure~\ref{cold_spectra}), so its contribution is incorporated when fitting the XMM-Newton data. Additionally, a power-law component is included in the ROSAT and XMM-Newton fits to model non-thermal emission, as its contribution extends to 0.1~keV. \cite{Bogdanov_2013} demonstrates that a power-law component is required to fit the XMM-Newton spectrum of PSR J0437--4715 adequately, and \cite{10.1093/mnras/stw2194} further constrain the power-law index using NuSTAR data spanning 2-20~keV (e.g.\ a power-law fit yields $\Gamma$~=~1.60$\pm$0.25 at 90\% confidence). Though pulsations are detected at 3.7$\sigma$ significance in the 2-20~keV band, they are only marginally detected (2.7$\sigma$ significance) in the 2-6~keV band. Since the hot-spot spectra with the hydrogen atmosphere can still contribute to the 2-6~keV band, and given the limited count statistics of the XMM-Newton data, the power-law non-thermal emission is treated as a phase-invariant component in the joint fit. 

The prior configurations and MultiNest settings are nearly identical to those described in Section~\ref{sec3.1} and \ref{sec3.2}. Owing to the limited number of counts in individual phase-energy bins, the simplest hot-spot geometric model is adopted: two spherical caps with uniform temperature distributions (ST-U). The adequacy of this model is discussed in Section~\ref{sec4.3}. Given the low signal-to-noise ratio of the high-energy tail in the XMM-Newton data, a Gaussian prior PDF is assigned to the power-law index and a uniform prior PDF to the normalization over the range (0.001-10)~$\times$10$^{-5}$~ph~keV$^{-1}$~cm$^{-2}$~s$^{-1}$. The total log-likelihood function for the joint multi-instrument fit is defined as,
\begin{equation}
\ln \mathcal{L}_\textup{total} = \ln \mathcal{L}_\textup{HST} + \ln \mathcal{L}_\textup{ROSAT} + \ln \mathcal{L}_\textup{XMM} \,,
\end{equation}
where the total log-likelihood takes the model parameters as input and returns a value to the MultiNest sampler. For the XMM-Newton EPIC-pn data, the energy threshold is set to 0.3~keV (lower than the 0.5~keV threshold recommended by the XMM-Newton Science Support Center) to maximize the utilization of available observation data. To account for potential instrumental artifacts, a background-marginalized likelihood function is adopted for the EPIC-pn data, following the approach described in Section~\ref{sec3.2}. The lower and upper bounds for the background constraints are defined as max(0,\{$B_\textup{XMM}$\}-3$\times$\{$\sigma_\textup{XMM}$\}) and (\{$B_\textup{XMM}$\}+3$\times$\{$\sigma_\textup{XMM}$\}) (for energies 0.3-3.0~keV), respectively. It should be noted that the imaging capability of XMM-Newton---enabled by its Wolter-I type mirrors and Charge-Coupled Device (CCD) detector---mitigates source contamination of the nearby bright AGN and allows more reliable instrumental background determination compared to NICER. 

\subsubsection{Inferred results}

\begin{table*}[htbp]
\centering
\caption{Summary of Bayesian parameter inference results for PSR J0437--4715, adopting the ST-U hot-spot geometric model.}
\label{finaltable}

\resizebox{\textwidth}{!}{
\begin{tabular}{lcccccccc}
\hline
			&&\multicolumn{2}{c}{{HST+ROSAT+XMM}} &\multicolumn{4}{c}{{HST+ROSAT+XMM+rPPA}} \\ 
\cline{5-8}   
			&&& &\multicolumn{2}{c}{{mode 1}} &\multicolumn{2}{c}{{mode 2}} \\ 
Parameter		&Prior &$\widehat{\textup{CI}}_{68\%}$ & Best-fit &$\widehat{\textup{CI}}_{68\%}$ & Best-fit&$\widehat{\textup{CI}}_{68\%}$ & Best-fit \\
\hline
$F_0$ (Hz)		 							
&173.69, fixed					& -	& -& -& -& -& - \\

$M$ (\(\textup{M}_\odot\)) 					
&$M\sim$~Gaussian(1.418,0.044)			& 1.48$\pm$0.02			 &1.50
						                  & 1.36$\pm$0.03		   &1.37
						                  & 1.38$\pm$0.03		   &1.40\\

$R_\textup{eq}$ (km) 						
&$R_\textup{eq}\sim$~Uniform(9, 17)		& 10.74$_{-0.68}^{+0.52}$    &11.26
						                  & 13.12$\pm$0.34        	   &13.33
						                  & 13.25$_{-0.35}^{+0.34}$      &13.51\\

$\cos(i)$ 					
&$\cos(i)\sim$~Gaussian(-0.7373,0.0002)		& -0.7373$\pm$0.0002		&-0.7372
						                      & -0.7373$\pm$0.0002		&-0.7371
						                      & -0.7373$\pm$0.0002		&-0.7373\\

$D$ (pc) 						
&$D\sim$~Uniform(156.979, 156.981)		& 156.980$\pm$0.001		&156.981
						                  & 156.980$\pm$0.001		&156.980
						                  & 156.980$\pm$0.001		&156.981\\

$N_\textup{H}$ (10$^{20}$ cm$^{-2}$)			
&$N_\textup{H}\sim$~Uniform(0.004, 2)		& 0.99$\pm$0.10			&0.91
						& 0.83$\pm$0.10			&0.71
						& 0.77$\pm$0.10			&0.78\\

$E(B-V)$ 					
&$E(B-V)\sim$~Gaussian(0.005,0.003) 		& 0.004$\pm$0.002		&0.005
						& 0.005$_{-0.003}^{+0.002}$		&0.005
						& 0.006$\pm$0.003		&0.009\\
&{truncated between 0.00001 and 0.07}& && \\

\hline
$\alpha_\textup{XMM}$ 				
&$\alpha_\textup{XMM}\sim$~Gaussian(1, 0.05)	& 1.00$\pm$0.04			&1.01
						& 1.03$\pm$0.04			&1.06
						& 1.04$\pm$0.04			&1.11\\

$\alpha_\textup{ROSAT}$ 				
&$\alpha_\textup{ROSAT}\sim$~Gaussian(1, 0.05)	& 0.98$\pm$0.04			&0.91
						& 0.97$\pm$0.04           &0.92
						& 0.97$\pm$0.04			&0.98\\

\hline
$\theta_\textup{p}$ (deg) 			
&$\cos(\theta_\textup{p})\sim$~Uniform(-1, 1)	& 17.16$_{-3.82}^{+4.52}$		&14.07
                        &-&- 
                        &-&-\\				
&or $\theta_\textup{p}\sim$~Gaussian(143, 9)	&-&-
						& 129.07$_{-2.30}^{+2.34}$		&127.68
						& 132.19$_{-2.20}^{+2.52}$		&130.30\\				

$\Delta\theta_\textup{p}$ (deg) 		
&$\Delta\theta_\textup{p}\sim$~Uniform(0.01, 45)& 20.28$_{-6.22}^{+4.57}$		&23.61
						& 3.02$_{-0.17}^{+0.15}$			&2.94
						& 3.08$_{-0.17}^{+0.15}$			&2.93\\

$kT_\textup{eff,p}$ (keV)			
&$kT_\textup{eff,p}\sim$~Uniform(0.011, 0.3)	& 0.106$_{-0.004}^{+0.005}$		&0.103
						& 0.098$\pm$0.002		&0.097
						& 0.097$\pm$0.002			&0.096\\

$\phi_\textup{p}$				
&$\phi_\textup{p}\sim$~Uniform(a+0.4, a+0.7)	& a+0.52$\pm$0.01		&a+0.51
						& a+0.52$\pm$0.01		&a+0.51
						& a+0.52$\pm$0.01		&a+0.53\\

$\theta_\textup{s}$ (deg) 			
&$\cos(\theta_\textup{s})\sim$~Uniform(-1, 1)	& 160.12$_{-2.84}^{+3.27}$		&158.83
						& 84.26$_{-6.18}^{+5.42}$		&80.08
						& 9.57$_{-1.20}^{+0.93}$		&9.36\\
      
$\Delta\theta_\textup{s}$ (deg)	 		
&$\Delta\theta_\textup{s}\sim$~Uniform(0.01, 45)& 2.52$\pm$0.16			&2.45
						& 1.63$_{-0.18}^{+0.15}$			&1.71
						& 28.82$_{-1.29}^{+1.43}$		&29.20\\

$kT_\textup{eff,s}$ (keV)	 		
&$kT_\textup{eff,s}\sim$~Uniform(0.011, 0.3)	& 0.113$\pm$0.005		&0.108
						& 0.114$\pm$0.005		&0.111
						& 0.096$\pm$0.003		&0.094\\

$\phi_\textup{s}$	 			
&$\phi_\textup{s}\sim$~Uniform(a-0.1, a+0.3)	& a+0.23$_{-0.03}^{+0.04}$		&a+0.21
						& a+0.08$\pm$0.01		&a+0.08
						& a+0.05$\pm$0.01		&a+0.05\\

\hline
$T_\textup{eff,bulk}$(10$^{5}$ K) 		
&$T_\textup{eff,bulk}\sim$~Uniform(1.5, 3.5)	& 2.59$_{-0.10}^{+0.11}$			&2.57
						& 2.19$_{-0.08}^{+0.10}$			&2.13
						& 2.19$_{-0.07}^{+0.11}$			&2.13\\

\hline
$\Gamma_\textup{PL}$ 				
&$\Gamma_\textup{PL}\sim$~Gaussian(1.60, 0.25) 	& 1.70$\pm$0.22 		&1.91
						& 2.26$_{-0.11}^{+0.15}$ 		&2.34
						& 2.12$_{-0.15}^{+0.18}$			&2.30\\

$N_\textup{PL}$\tiny{(10$^{-5}$ ph keV$^{-1}$ cm$^{-2}$ s$^{-1}$)}	
&$N_\textup{PL}\sim$~Uniform(0.001, 10) 	& 2.94$_{-0.70}^{+0.72}$			&3.94
						& 6.99$\pm$1.02			            &7.90
						& 5.80$_{-1.09}^{+0.97}$			&6.30\\
\hline
\end{tabular}
}
\end{table*}

The multi-wavelength spectra and energy-resolved pulse profiles of PSR J0437--4715 are simultaneously fit across the FUV and X-ray bands, using observations from HST, ROSAT, and XMM-Newton. The adopted priors, inferred parameters with their 68\% CIs, and best-fit values are summarized in Table~\ref{finaltable}. The data products and Python scripts required to reproduce the results and figures presented in this work are available in the Zenodo repository\footnote{\url{https://doi.org/10.5281/zenodo.18092208}}. The fit quality with the best-fit values is assessed by plotting the residual distributions; the absence of discernible structures therein confirms that the model provides a good description of the observational data.

The inferred neutron star mass, inclination angle, and distance are predominantly constrained by the tightly bound prior PDFs derived from pulsar radio timing measurements~\citep{Reardon_2024}. The inferred equatorial circumferential radius of 10.74$_{-0.68}^{+0.52}$~km is consistent with the results of \cite{choudhury2024nicer} at the 68\% CI level, albeit with a median value that is 0.62~km smaller. The inferred column density of (0.99$\pm$0.10)~$\times$10$^{20}$~cm$^{-2}$ is consistent with the Dispersion Measure (DM) and extinction measurements via empirical relations, in contrast to the relatively lower values of (0.05$_{-0.04}^{+0.68}$)~$\times$10$^{20}$~cm$^{-2}$ inferred from NICER-only data~\citep{choudhury2024nicer}. The interstellar dust extinction of 0.004$\pm$0.002 is in good agreement with previous measurements~\citep{2018AA...616A.132L,gaia}. The inferred effective-area scaling factors for ROSAT and XMM-Newton, centered at $\approx$1, are consistent with their respective Gaussian prior PDFs. The inferred hot-spot effective temperatures are (1.23$_{-0.05}^{+0.06}$) and (1.31$\pm$0.06)~$\times$10$^6$~K, with an effective bulk surface temperature of (2.59$_{-0.10}^{+0.11}$)~$\times$10$^5$~K; these values are consistent with other estimates and analyses~\citep{Durant_2012,Bogdanov_2013,10.1093/mnras/stz2941,choudhury2024nicer}. Additionally, the inferred power-law index of 1.70$\pm$0.22 and norm of (2.94$_{-0.70}^{+0.72}$)~$\times$10$^{-5}$~ph~keV$^{-1}$~cm$^{-2}$~s$^{-1}$ are consistent with the spectral fitting results obtained using XMM-Newton data spanning 0.1-10.0~keV~\citep{Bogdanov_2013} and NuSTAR data extending up to 20~keV~\citep{10.1093/mnras/stw2194}. 

However, the inferred hot-spot geometry consists of two spherical caps located at nearly symmetric colatitudes relative to the equator, with a rotational phase offset of 0.71. The primary hot spot---corresponding to the main pulse peak---is situated in the northern hemisphere and lies far from the observer's line of sight. The secondary hot spot is positioned close to the south pole. This geometric configuration is inconsistent with radio observations, both in terms of the pulse profile morphology and polarization properties. Radio pulse profiles indicate that pulsar emission is seen only from one magnetic pole~\citep{Bhat_2014}. Fitting the Rotation Vector Model (RVM) to the position angle distribution yields geometric parameters of $\alpha$~=~145$^{\circ}$ and $\zeta$~=~140$^{\circ}$~\citep{1995ApJ...441L..65M}, where $\alpha$ denotes the angle between the magnetic and rotation axes, and $\zeta$ denotes the angle between the observer's line of sight and rotation axes. Given that the absolute phase of the main pulse peak in the radio band aligns with that in the X-ray band~\citep{Bogdanov_2013}, the radio and X-ray radiation are expected to originate from spatially coincident or closely adjacent regions---most likely from or close to the polar cap. Thus, while Bayesian inference yields a statistically preferred solution that successfully reproduces the UV and X-ray data, it conflicts with the radio measurements. The confidence in the radio measurements suggests that this solution likely does not reflect the actual physical emission geometry.

\subsection{HST+ROSAT+XMM+rPPA}
\label{sec4.3}
The RVM traces the global dipole magnetic field geometry from extended radio emission in the magnetosphere, whereas the PPM constrains the surface hot-spot position from thermal emission on the neutron star surface. Radio emission originates from regions close to the stellar surface and is sensitive to potential multipolar magnetic field components; thus, the hot-spot location may be decoupled from the radio viewing geometry when magnetic field lines are strongly distorted. For PSR J0437$-$4715, the absolute phase of the main pulse peak in the radio band aligns with that in the X-ray band. Accordingly, an informative prior derived from radio polarization position angle (rPPA) measurements is applied to constrain the hot-spot geometry in the PPM framework, i.e.\ HST+ROSAT+XMM+rPPA.

The rPPA fit is re-performed on the dataset presented in \cite{1995ApJ...441L..65M} using the RVM. The presence of orthogonal polarization modes (OPMs) is neglected in the present analysis, which results in a poor fitting performance as the observed rPPA deviates significantly from the ideal RVM S-shaped curve. The formal parameter uncertainties are scaled by a factor of $\sqrt{\chi^2_{\rm red}}$~=~4 to account for this reduced fitting quality. The fitted magnetic inclination angle $\alpha$ is 143$\pm$9$^\circ$, and the observer line-of-sight angle $\zeta$ is 130$\pm$13$^\circ$. These results are consistent with those reported in \cite{petri2026multi}, i.e.\ $\alpha$~=~132$^{\circ}$-149$^{\circ}$ at different wavelengths of 10~cm, 20~cm, and 50~cm, where the OPM jumps are identified with a sufficient degree of linear polarization ($\ge$30\%). Setting a high threshold on the linear polarization degree improves the RVM recovery~\citep{10.1093/mnras/stae1175}, which is out of the scope of the current work. Consequently, a Gaussian prior PDF is assigned to the central colatitude of the primary hot spot, with a mean of 143$^\circ$ and a standard deviation of 9$^\circ$. A preliminary Bayesian inference run reveals multi-modal structures in the posterior parameter space. To ensure comprehensive sampling of this complex parameter space, the number of live points is increased from 1000 to 2000, and the multi-mode sampling option is enabled---with the maximum number of supported modes set to 5.

\begin{figure}[h]
\begin{center}
\includegraphics[width=0.7\linewidth]{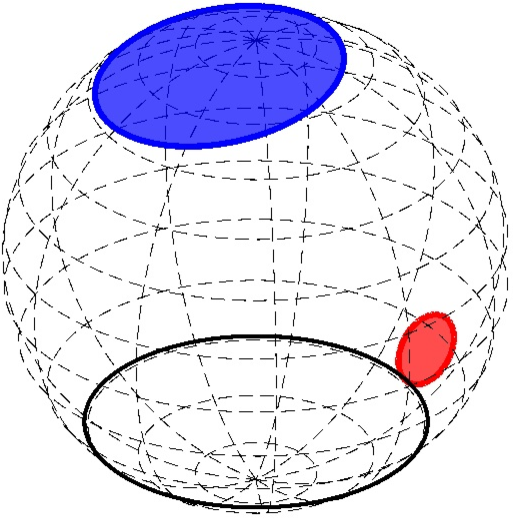}
\end{center}
\caption{Schematic illustration of the two circular hot-spot geometry inferred from the best-fit parameters of the second posterior mode. The primary hot spot is shown in red, and the secondary hot spot in blue. The observer's line of sight is indicated by the black line, corresponding to a colatitude of $137.5^\circ$. Exact geometric parameters are listed in Table~\ref{finaltable}.}
\label{geometry}
\end{figure}

\begin{figure}[h]
\begin{center}
\includegraphics[width=0.85\linewidth]{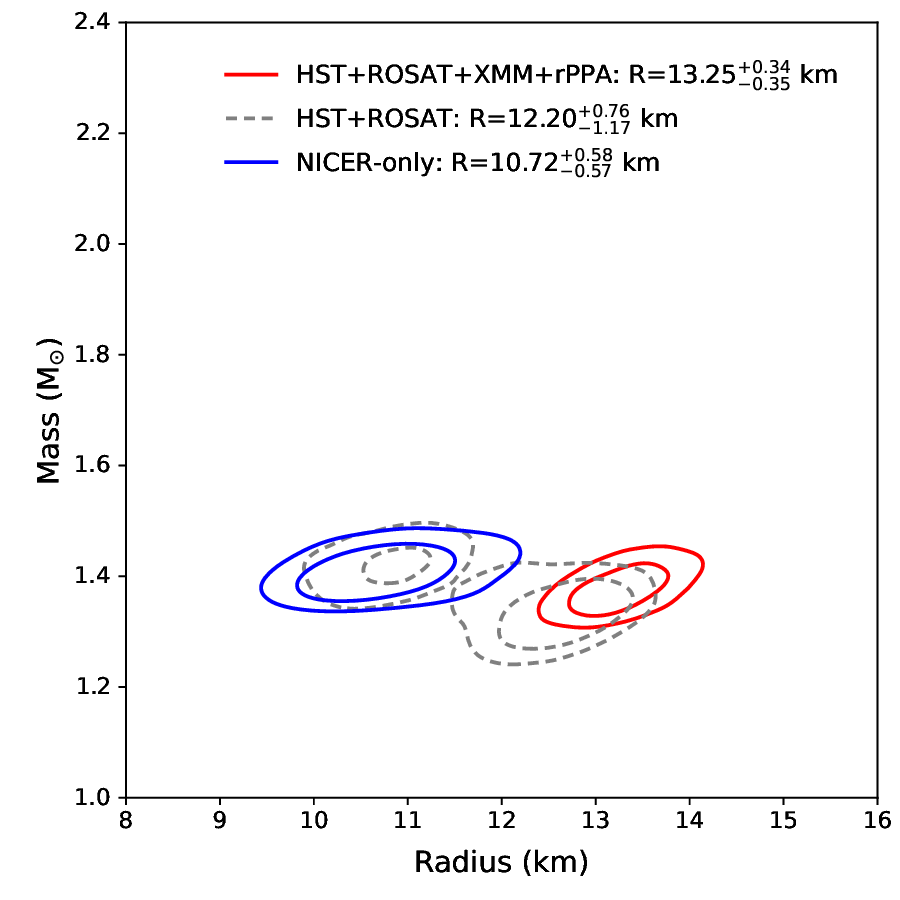}
\end{center}
\caption{Two-dimensional marginalized posterior PDFs of neutron star mass and radius, comparing results from different datasets and fitting methodologies employed in this work. The contours represent the 68\% and 95\% credible regions.}
\label{MRfinal}
\end{figure}

The adopted prior PDFs and results of Bayesian parameter estimation are summarized in Table~\ref{finaltable}. MultiNest identifies two distinct posterior modes. The inferred geometry of the first mode consists of two hot spots, with the colatitude of the secondary hot spot near the equator. By contrast, the second mode yields a more physically compelling solution: the primary hot spot is situated at a colatitude of $\approx$130$^{\circ}$, while the secondary hot spot lies at a colatitude of $\approx$9$^{\circ}$, close to the north pole (see Figure~\ref{geometry}). Emission from the secondary hot spot is not expected to be detectable in the radio band, and the magnetic inclination angle derived from the primary hot spot's colatitude remains consistent with the position angle distribution via the RVM. Additionally, at a qualitative level, the inferred hot-spot geometry is consistent with that derived from NICER-only data, in terms of both hot-spot position and size (see Table~\ref{niceronlytable}). The log-evidence value of the second mode is 1.0 unit higher than that of the first mode, indicating that this solution is statistically preferred and thus presented as the headline result of this work. The two-dimensional marginalized posterior PDFs of neutron star masses and radii for this headline result are plotted in Figure~\ref{MRfinal}, alongside the Bayesian parameter estimates from Section~\ref{sec3.1} and \ref{sec3.2}. Compared to the NICER-only analysis, the radio-consistent radius posteriors derived from the HST+ROSAT+XMM+rPPA dataset shift toward larger values. Compared to the spectral fit of the HST+ROSAT dataset, it yields tighter constraints on the neutron star radius.    

\begin{figure}[h]
\begin{center}
\begin{minipage}{0.49\linewidth}
\includegraphics[width=1\linewidth]{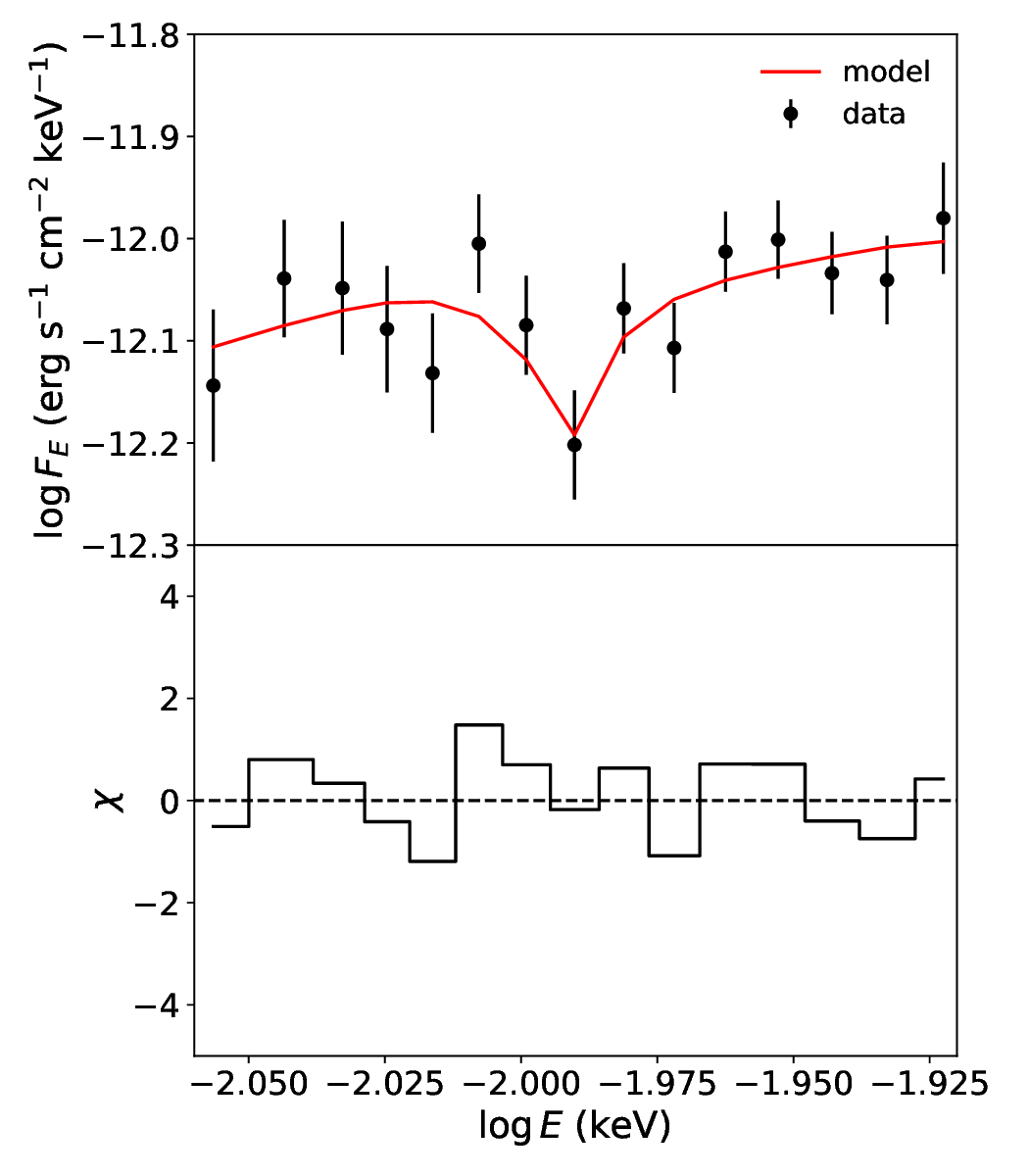}
\end{minipage}
\begin{minipage}{0.49\linewidth}
\includegraphics[width=1\linewidth]{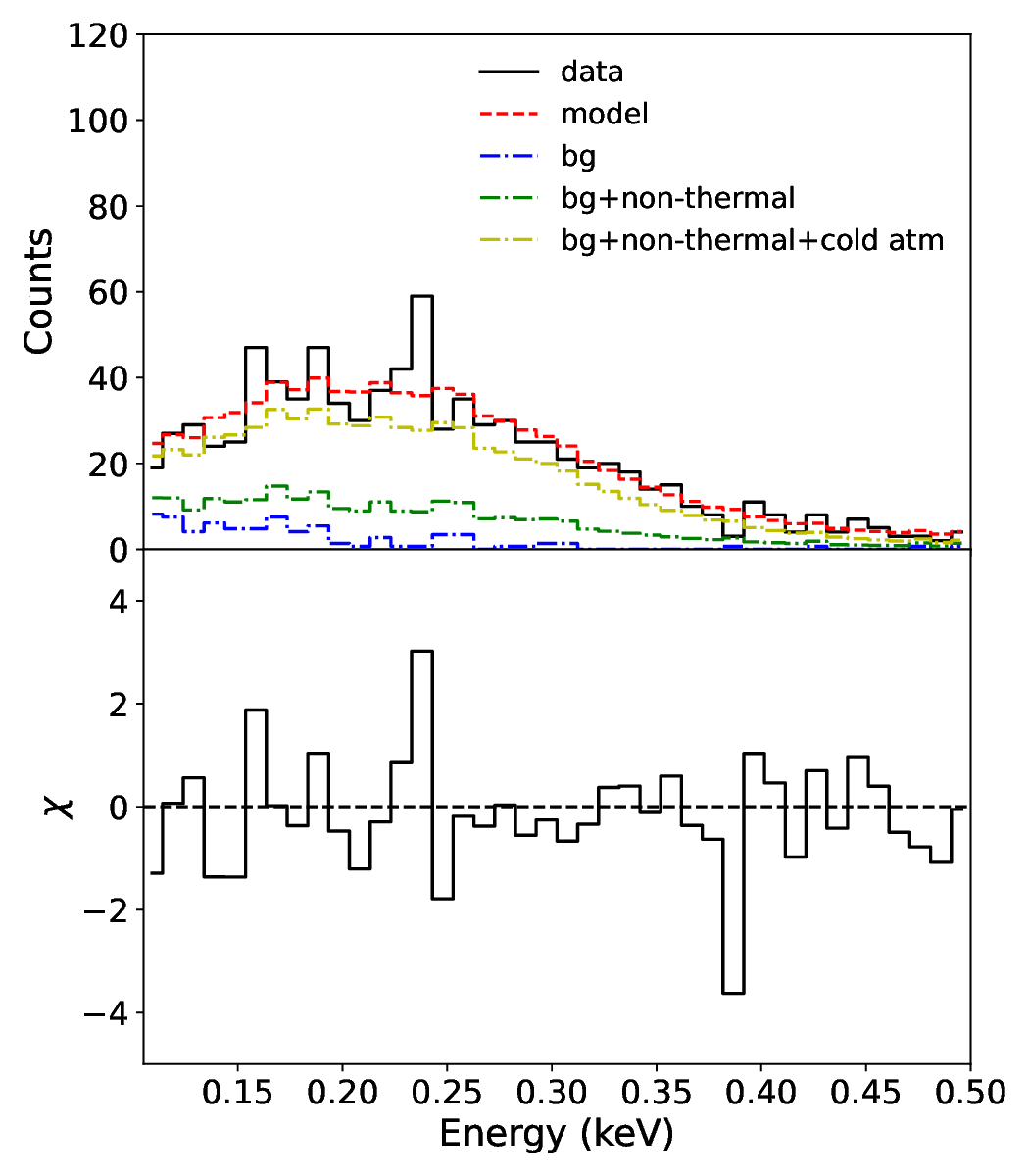}
\end{minipage}
\end{center}
\caption{Left panel: Comparison of non-redshifted spectral flux distributions and fit residuals between HST FUV data and the best-fit cold atmosphere model. Right panel: Comparison of energy spectra and fit residuals between ROSAT soft X-ray data and the best-fit multi-component model.}
\label{check1}
\end{figure}

\begin{figure}[h]
\begin{center}
\begin{minipage}{0.49\linewidth}
\includegraphics[width=1\linewidth]{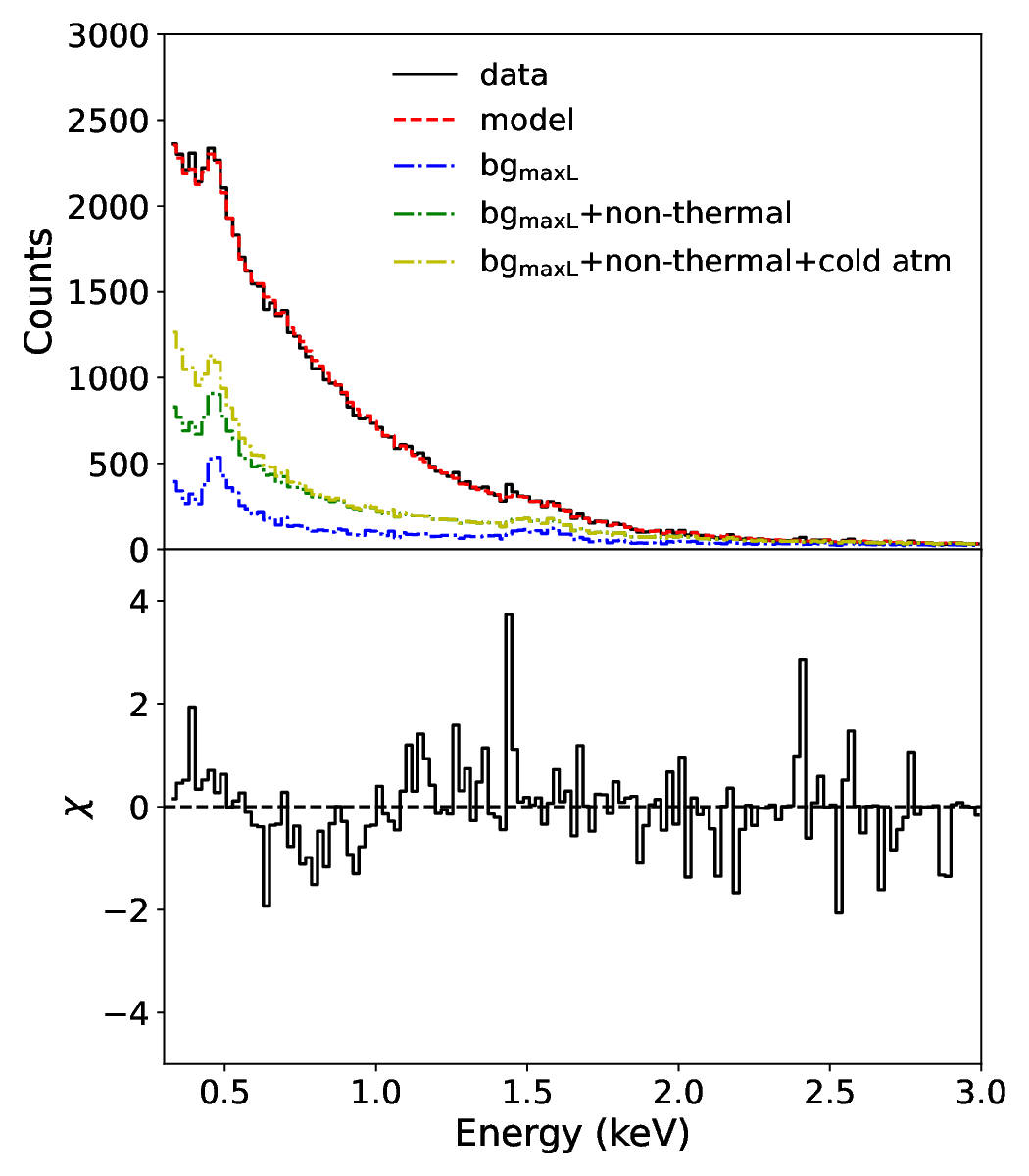}
\end{minipage}
\begin{minipage}{0.49\linewidth}
\includegraphics[width=1\linewidth]{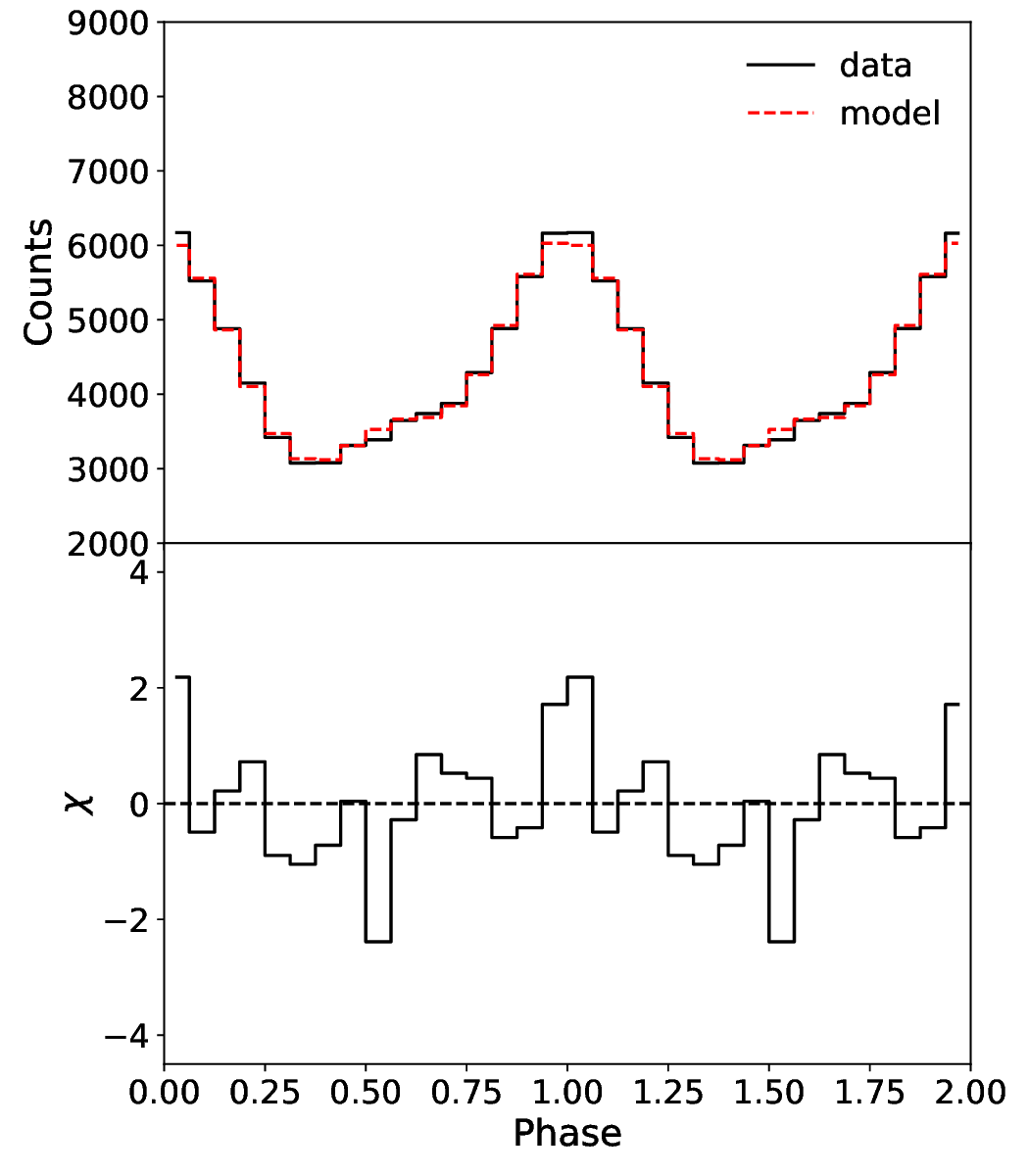}
\end{minipage}
\end{center}
\caption{Left panel: Comparison of total energy spectra and fit residuals between XMM-Newton EPIC-pn data and the best-fit multi-component model. Right panel: Comparison of energy-integrated pulse profiles between XMM-Newton EPIC-pn data and the best-fit model.}
\label{check2}
\end{figure}

\begin{figure}[h]
\begin{center}
\includegraphics[width=0.8\linewidth]{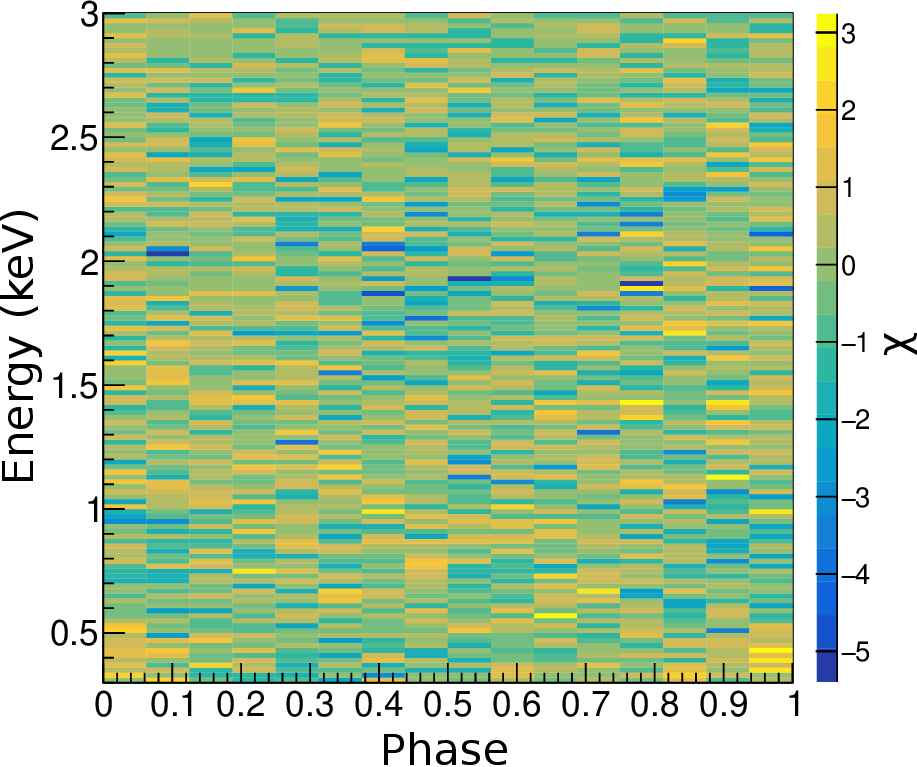}
\end{center}
\caption{Fit residuals of energy-resolved pulse profiles between XMM-Newton EPIC-pn data and the best-fit model.}
\label{check3}
\end{figure}

To assess the fit quality of the headline model, residual distributions calculated using the best-fit values are presented in Figure~\ref{check1}, \ref{check2}, and \ref{check3}. No apparent structure is present, suggesting that the model can reproduce the data. In the left panel of Figure~\ref{check1}, hot-spot contributions are excluded from the FUV band analysis; the inferred geometry justifies this assumption, as the primary hot spot is compact and the secondary hot spot is obscured from the observer. The right panel of Figure~\ref{check1} and Figure~\ref{check2} demonstrates that, in the X-ray band below $\approx$0.7~keV, contributions from both cold thermal and non-thermal emission are non-negligible relative to hot thermal emission. Furthermore, non-thermal emission dominates the high-energy tail of the XMM-Newton spectrum. Including non-thermal emission in the model tends to reduce the inferred hot-spot temperature. Figure~\ref{check3} displays the residual distribution of the fit to the energy-resolved pulse profiles. Owing to limited photon statistics, a small number of phase-energy bins yield relatively large negative $\chi$ values, with a minimum of -5.4. Furthermore, the fit quality is assessed by introducing additional complexity to the hot-spot geometry, specifically a model of Concentric Single-temperature Regions with Unshared parameters (CST-U). The inferred posteriors for the angular radius ratio of the two spots remain poorly constrained, while all other parameters exhibit no significant deviation from the headline results. This indicates that the log-likelihood function is not sensitive to more complex hot-spot geometries given the limited statistical quality of the XMM-Newton data.

It is important to note that the joint spectral and timing fit utilizing HST, ROSAT, and XMM-Newton data is subject to several limitations that may introduce biases into the inferred hot-spot geometry and neutron star $M$-$R$ measurements for PSR J0437--4715. First, statistical and systematic uncertainties in flux measurements across the FUV and soft X-ray bands ($<$0.3~keV), combined with a lack of observational data bridging the FUV and soft X-ray regimes, constrain the quality of the fitting results. Second, the limited statistics of individual phase-energy bins in the XMM-Newton dataset restrict the model fit performance; in particular, the low signal-to-noise ratio of the high-energy tail leads to an inferred non-thermal power-law spectrum that is softer than that derived from NuSTAR data. Third, the exclusion of pulsations in the non-thermal emission may introduce additional biases. Finally, tighter constraints on hot-spot geometry are required to mitigate model degeneracies and associated systematic biases. To overcome these limitations, the enhanced X-ray Timing and Polarimetry (eXTP) mission represents a promising solution, which has been fully approved for launch in 2030~\citep{Zhang2025}. eXTP lists determining the equation of state of matter at supra-nuclear density as one of its core scientific objectives~\citep{Li2025}. One unit of the Spectroscopy Focusing Array (SFA) is equipped with a Wolter-I focusing mirror and a pn-CCD focal-plane camera (designated SFA-I, where "I" denotes imaging), achieving an effective area $>$820~cm$^2$ at 1.5~keV. Mega-second-scale observations are planned for several millisecond pulsars. These observations will enable tighter constraints on neutron star radii when analyzed with the methodology presented in this work.

\subsection{Impact of modulated non-thermal emission}
The non-thermal emission component is modeled as a phase-invariant power-law distribution. This choice is motivated by two key factors: first, it has been widely employed in previous studies to reproduce the observed energy spectra~\citep{Durant_2012,Bogdanov_2013,10.1093/mnras/stw2194}; second, its mathematical simplicity facilitates its straightforward integration into the PPM framework. The impact of possible pulsations in this non-thermal component on the headline results is quantitatively assessed in this section.

\begin{figure}[h]
\begin{center}
\includegraphics[width=0.7\linewidth]{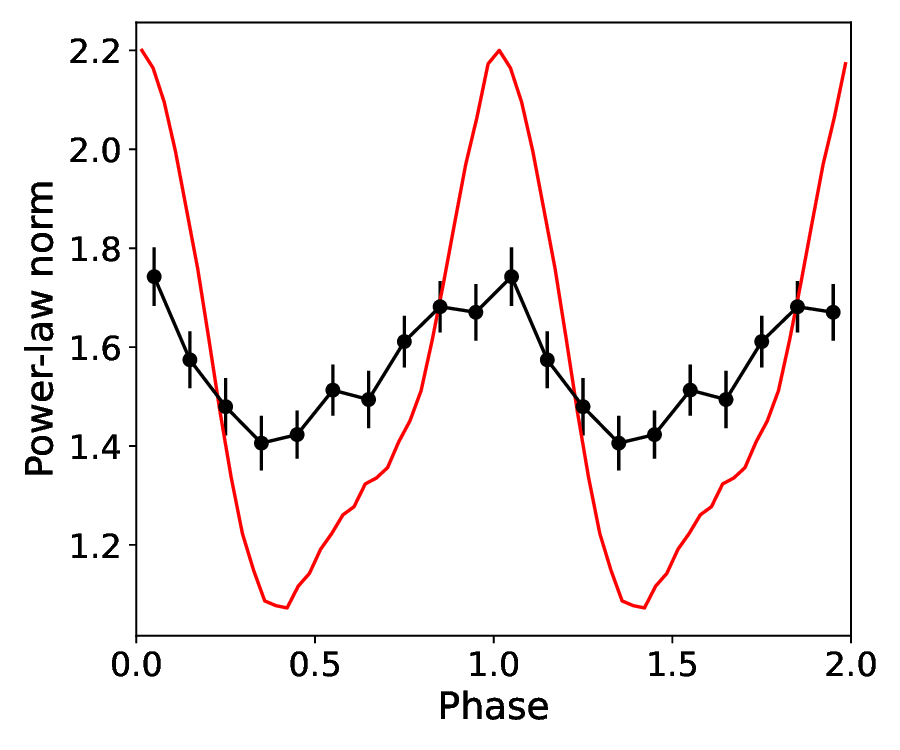}
\end{center}
\caption{Distribution of the power-law normalization as a function of rotational phase, binned into 10 intervals. Two rotational cycles are displayed for clarity. The energy-integrated pulse profile is overplotted in red for reference.}
\label{PL_norm}
\end{figure}

Phase-resolved energy spectra (10 phase intervals) are extracted from the NICER data. Each spectrum is modeled over the 0.3-3.0 keV energy range using a composite model consisting of three blackbody components (3BB) plus a power-law component (PL)~\citep{Zheng2026prep}. Following \cite{10.1093/mnras/stw2194}, the neutral hydrogen column density is fixed at 0.24$\times$10$^{20}$ cm$^{-2}$, and the photon index of the power-law component is fixed to $\Gamma$~=~1.65 for each spectral fit. The best-fit power-law normalization as a function of rotational phase is extracted (see Figure~\ref{PL_norm}), from which a pulsed fraction of $\mathcal{F}$~=~0.11$\pm$0.02 is derived.

To quantify the impact of non-thermal pulsations on the inference, a set of synthetic observational data is generated incorporating a power-law contribution modulated by the estimated pulsed fraction. Bayesian inference is then performed on this synthetic dataset using a phase-invariant power-law model. A comparison of the input parameters and the inferred posteriors for the synthetic data is summarized in Table~\ref{PLtable}. The results demonstrate that, for a pulsed fraction of $\mathcal{F} = 0.11$ in this non-thermal emission, the impact on the inferred neutron star radius, mass, and hot-spot geometry is statistically insignificant, i.e.\ the input value lies almost within the 68\% CIs of the inferred posteriors.

\begin{table}[htp]
\centering
\caption{Summary of input parameters and inferred results for synthetic data with a modulated power-law contribution.}
\resizebox{\linewidth}{!}{
\begin{tabular}{lccc}
\hline
Parameter   &Input &$\widehat{\textup{CI}}_{68\%}$ & Best-fit \\
\hline
$F_0$ (Hz)		 							
&173.69, fixed					& -	& - \\

$M$ (\(\textup{M}_\odot\)) 					
&1.4        & 1.36$\pm$0.03             &1.31\\

$R_\textup{eq}$ (km) 						
&13.4       & 13.04$_{-0.31}^{+0.32}$            &12.98\\

$\cos(i)$ 					
&-0.7373	& -0.7373$\pm$0.0002		&-0.7374\\

$D$ (pc) 						
&156.98	    & 156.980$\pm$0.001		   &156.980\\

$N_\textup{H}$ (10$^{20}$ cm$^{-2}$)			
&0.8		& 0.78$\pm$0.09			&0.76\\

$E(B-V)$ 					
&0.005 		& 0.006$\pm$0.003		&0.002\\

\hline
$\alpha_\textup{XMM}$   
&1          & 1.01$\pm$0.04			&1.01\\

$\alpha_\textup{ROSAT}$ 				
&1          & 0.99$\pm$0.04			&1.00\\

\hline
$\theta_\textup{p}$ (deg) 			
&130        & 133.02$\pm$2.06		&132.64\\		
                        
$\Delta\theta_\textup{p}$ (deg) 		
&3          & 3.07$\pm$0.14		&3.10\\

$kT_\textup{eff,p}$ (keV)			
&0.1        & 0.102$\pm$0.001		&0.100\\

$\phi_\textup{p}$				
&a+0.52     & a+0.52$\pm$0.01		&a+0.52\\

$\theta_\textup{s}$ (deg) 			
&9          & 8.37$_{-0.79}^{+0.78}$		&7.53\\
      
$\Delta\theta_\textup{s}$ (deg)	 		
&29         & 29.10$_{-1.25}^{+1.32}$			&30.39\\

$kT_\textup{eff,s}$ (keV)	 		
&0.1        & 0.104$\pm$0.003			&0.105\\

$\phi_\textup{s}$	 			
&a+0.05     & a+0.05$\pm$0.01           		&a+0.05\\

\hline
$T_\textup{eff,bulk}$(10$^{5}$ K) 		
&2.14       & 2.24$_{-0.04}^{+0.05}$		&2.25\\

\hline
$\Gamma_\textup{PL}$
&2.2 	& 1.90$\pm$0.23 		&2.13\\

$N_\textup{PL}$\tiny{(10$^{-5}$ ph keV$^{-1}$ cm$^{-2}$ s$^{-1}$)}	
&6.5(phase-averaged) 	& 2.37$\pm$0.66			&3.58\\

\hline
\end{tabular}
}
\label{PLtable}
\end{table}

\section{Conclusions}
\label{sec5}

This work presents a comprehensive multi-wavelength analysis of the nearby and bright millisecond pulsar PSR J0437--4715, integrating HST FUV, ROSAT soft X-ray, and XMM-Newton X-ray observations to characterize its emission components, constrain its neutron star radius, and infer the geometric properties of its hot spots via Bayesian inference.
 
The analysis yields key results from the joint fit of multi-wavelength spectra and energy-resolved pulse profiles. First, it confirms multi-component emission in this pulsar's broadband spectrum, with comparable contributions from cold thermal, hot thermal, and non-thermal emission in the low-energy tail, and non-thermal emission dominating the high-energy tail in the XMM-Newton data. Second, it identifies a statistically viable hot-spot geometry---specifically, the primary hot spot situated at a colatitude of $\approx$130$^{\circ}$ and the secondary hot spot at a colatitude of $\approx$9$^{\circ}$---consistent with radio observations and qualitatively aligned with NICER-only results. Finally, the solution of HST+ROSAT+XMM+rPPA provides tighter radius constraints (13.25$_{-0.35}^{+0.34}$~km, 68\% CI) than HST+ROSAT fits and shifts the radius posterior distribution to larger values relative to NICER-only results. 

The analysis also highlights several limitations: statistical and systematic uncertainties in FUV and soft X-ray ($<$0.3 keV) flux measurements, limited photon statistics in XMM-Newton phase-energy bins (particularly in the high-energy tail), potential biases from excluding non-thermal pulsations, and model degeneracies in hot-spot geometry. These limitations may impact the precision of inferred $M$-$R$ relations and geometric parameters. 

Future observations from the eXTP mission---equipped with large-effective-area Wolter-I mirrors and planned mega-second-scale exposures of millisecond pulsars---will mitigate these limitations. eXTP's capabilities will enable tighter constraints on neutron star radii and EOS via the methodology established in this work, advancing our understanding of dense matter at supranuclear densities. Additionally, improved background modeling, along with possibly more refined hot-spot geometry priors, will further reduce systematic biases in future analyses of PSR J0437--4715.

\begin{acknowledgments}
This research utilized data and software from HEASARC, provided by NASA's Goddard Space Flight Center and observation data obtained with HST, ROSAT, XMM-Newton, and NICER. This work is supported by China's Space Origins Exploration Program. The authors thank supports from Grant No. E4298RU8 of Institute of High Energy Physics, CAS. Mingyu Ge thanks the support of National Natural Science Foundation of China (Nos. 12373051 and 12333007). Ang Li thanks the support of National Natural Science Foundation of China (No. 12273078). Weiwei Xu thanks the support of National Nature Science Foundation of China (Nos. 11988101, 12022306, 12203063, and 12333007), the support by National Key R\&D Program of China No. 2022YFF0503403, the support from the Ministry of Science and Technology of China (Nos. 2020SKA0110100), the science research grants from the China Manned Space Project (Nos. CMS-CSST-2025-A03, CMS-CSST-2021-B01, and CMS-CSST-2021-A01), CAS Project for Young Scientists in Basic Research (No. YSBR-062), and the support from K.C. Wong Education Foundation. We would like to thank Pengfei Zhang for assistance with the data analysis.
\end{acknowledgments}

\bibliography{mybib}{}

@Article{Zhang2025,
author={Zhang, Shuang-Nan
and Santangelo, Andrea
and Xu, Yupeng
and Feng, Hua
and Lu, Fangjun
and Chen, Yong
and Ge, Mingyu
and Nandra, Kirpal
and Wu, Xin
and Feroci, Marco
and Hernanz, Margarita
and Liu, Congzhan
and He, Huilin
and Wang, Yusa
and Jiang, Weichun
and Cui, Weiwei
and Yang, Yanji
and Wang, Juan
and Li, Wei
and Li, Hong
and Du, Yuanyuan
and Liu, Xiaohua
and Meng, Bin
and Wen, Xiangyang
and Zhang, Aimei
and Ma, Jia
and Li, Maoshun
and Li, Gang
and Qi, Liqiang
and Sun, Jianchao
and Luo, Tao
and Liu, Hongwei
and Liu, Xiaojing
and Zhang, Fan
and Luo, Laidan
and Zhu, Yuxuan
and Zhao, Zijian
and Sun, Liang
and Yang, Xiongtao
and Wu, Qiong
and Jiang, Jiechen
and Shi, Haoli
and Liu, Jiangtao
and Xu, Yanbing
and Yang, Sheng
and Zhang, Laiyu
and Han, Dawei
and Gao, Na
and Huo, Jia
and Zhang, Ziliang
and Wang, Hao
and Zhao, Xiaofan
and Wang, Shuo
and Li, Zhenjie
and Bao, Ziyu
and Liu, Yaoguang
and Wang, Ke
and Wang, Na
and Wang, Bo
and Wang, Langping
and Wang, Dianlong
and Ding, Fei
and Sheng, Lizhi
and Qiang, Pengfei
and Yan, Yongqing
and Liu, Yongan
and Wu, Zhenyu
and Liu, Yichen
and Chen, Hao
and Zhang, Yacong
and Liu, Hongbang
and Altmann, Alexander
and Bechteler, Thomas
and Burwitz, Vadim
and Fiorini, Carlo
and Friedrich, Peter
and Meidinger, Norbert
and Strecker, Rafael
and Baldini, Luca
and Bellazzini, Ronaldo
and Bonino, Raffaella
and Frass{\`a}, Andrea
and Latronico, Luca
and Maldera, Simone
and Manfreda, Alberto
and Minuti, Massimo
and Pesce-Rollins, Melissa
and Sgr{\`o}, Carmelo
and Tugliani, Stefano
and Pareschi, Giovanni
and Basso, Stefano
and Sironi, Giorgia
and Spiga, Daniele
and Tagliaferri, Gianpiero
and Tykhonov, Andrii
and Paltani, St{\`e}phane
and Bozzo, Enrico
and Tenzer, Christoph
and Bayer, J{\"o}rg
and Tuo, Youli
and Liu, Honghui
and Zhang, Yonghe
and Cai, Zhiming
and Liu, Huaqiu
and Chen, Wen
and Wang, Chunhong
and He, Tao
and Chen, Yehai
and Qiu, Chengbo
and Zhang, Ye
and Feng, Jianchao
and Zhu, Xiaofei
and Zhou, Heng
and Zheng, Shijie
and Song, Liming
and Wang, Jinzhou
and Jia, Shumei
and Jiang, Zewen
and Li, Xiaobo
and Zhao, Haisheng
and Guan, Ju
and Zhang, Juan
and Li, Chengkui
and Huang, Yue
and Liao, Jinyuan
and You, Yuan
and Zhang, Hongmei
and Wang, Wenshuai
and Wang, Shuang
and Ou, Ge
and Hu, Hao
and Shi, Jingyan
and Cui, Tao
and Jiang, Xiaowei
and Cheng, Yaodong
and Li, Haibo
and Xu, Yanjun
and Zane, Silvia
and Bambi, Cosimo
and Bu, Qingcui
and Dall'Osso, Simone
and Rosa, Alessandra De
and Gou, Lijun
and Guillot, Sebastien
and Ji, Long
and Li, Ang
and Mao, Jirong
and Patruno, Alessandro
and Stratta, Giulia
and Taverna, Roberto
and Tsygankov, Sergey
and Uttley, Phil
and Watts, Anna L.
and Wu, Xuefeng
and Xu, Renxin
and Yi, Shuxu
and Zhang, Guobao
and Zhang, Liang
and Zhao, Wen
and Zhou, Ping},
title={The enhanced X-ray Timing and Polarimetry mission---eXTP for launch in 2030},
journal={Science China Physics, Mechanics {\&} Astronomy},
year={2025},
month={Sep},
day={24},
volume={68},
number={11},
pages={119502},
issn={1869-1927},
doi={10.1007/s11433-025-2786-6},
url={https://doi.org/10.1007/s11433-025-2786-6}
}

@Article{Li2025,
author={Li, Ang
and Watts, Anna L.
and Zhang, Guobao
and Guillot, Sebastien
and Xu, Yanjun
and Santangelo, Andrea
and Zane, Silvia
and Feng, Hua
and Zhang, Shuang-Nan
and Ge, Mingyu
and Qi, Liqiang
and Salmi, Tuomo
and Dorsman, Bas
and Miao, Zhiqiang
and Tu, Zhonghao
and Cavecchi, Yuri
and Zhou, Xia
and Zheng, Xiaoping
and Wang, Weihua
and Cheng, Quan
and Liu, Xuezhi
and Wei, Yining
and Wang, Wei
and Xu, Yujing
and Weng, Shanshan
and Zhu, Weiwei
and Li, Zhaosheng
and Shao, Lijing
and Tuo, Youli
and Dohi, Akira
and Lyu, Ming
and Liu, Peng
and Yuan, Jianping
and Wang, Mingyang
and Zhang, Wenda
and Li, Zexi
and Tao, Lian
and Zhang, Liang
and Shen, Hong
and Provid{\^e}ncia, Constan{\c{c}}a
and Tolos, Laura
and Patruno, Alessandro
and Li, Li
and Liu, Guozhu
and Zhou, Kai
and Chen, Lie-Wen
and Fan, Yizhong
and Kajino, Toshitaka
and Lai, Dong
and Li, Xiangdong
and Meng, Jie
and Tang, Xiaodong
and Xiao, Zhigang
and Xiong, Shaolin
and Xu, Renxin
and Zhou, Shan-Gui
and Ballantyne, David R.
and Burgio, G. Fiorella
and Chenevez, J{\'e}r{\^o}me
and Choudhury, Devarshi
and Fantina, Anthea F.
and Galloway, Duncan K.
and Gulminelli, Francesca
and Hebeler, Kai
and Hoogkamer, Mariska
and Horvath, Jorge E.
and Kini, Yves
and Kurkela, Aleksi
and Linares, Manuel
and Margueron, J{\'e}r{\^o}me
and Mendes, Melissa
and Oertel, Micaela
and Papitto, Alessandro
and Poutanen, Juri
and Rea, Nanda
and Schwenk, Achim
and Song, Xin-Ying
and Svensson, Isak
and Tsang, David
and Vuorinen, Aleksi
and Andersson, Nils
and Miller, M. Coleman
and Rezzolla, Luciano
and Stone, Jirina R.
and Thomas, Anthony W.},
title={Dense matter in neutron stars with eXTP},
journal={Science China Physics, Mechanics {\&} Astronomy},
year={2025},
month={Sep},
day={18},
volume={68},
number={11},
pages={119503},
issn={1869-1927},
doi={10.1007/s11433-025-2761-4},
url={https://doi.org/10.1007/s11433-025-2761-4}
}

@article{jansen2001xmm,
  title={XMM-Newton observatory-I. The spacecraft and operations},
  author={Jansen, F and Lumb, D and Altieri, B and Clavel, J and Ehle, M and Erd, C and Gabriel, C and Guainazzi, M and Gondoin, P and Much, R and others},
  journal={Astronomy \& Astrophysics},
  volume={365},
  number={1},
  pages={L1--L6},
  year={2001},
  publisher={EDP Sciences}
}

@article{struder2001european,
  title={The European photon imaging camera on XMM-Newton: the pn-CCD camera},
  author={Str{\"u}der, L and Briel, U and Dennerl, K and Hartmann, R and Kendziorra, E and Meidinger, N and Pfeffermann, E and Reppin, C and Aschenbach, B and Bornemann, W and others},
  journal={Astronomy \& Astrophysics},
  volume={365},
  number={1},
  pages={L18--L26},
  year={2001},
  publisher={EDP Sciences}
}

@inproceedings{gendreau2016neutron,
  title={The neutron star interior composition explorer (NICER): design and development},
  author={Gendreau, Keith C and Arzoumanian, Zaven and Adkins, Phillip W and Albert, Cheryl L and Anders, John F and Aylward, Andrew T and Baker, Charles L and Balsamo, Erin R and Bamford, William A and Benegalrao, Suyog S and others},
  booktitle={Space telescopes and instrumentation 2016: Ultraviolet to gamma ray},
  volume={9905},
  pages={420--435},
  year={2016},
  organization={SPIE}
}

@article{TRUMPER1982241,
title = {The ROSAT mission},
journal = {Advances in Space Research},
volume = {2},
number = {4},
pages = {241-249},
year = {1982},
issn = {0273-1177},
doi = {https://doi.org/10.1016/0273-1177(82)90070-9},
url = {https://www.sciencedirect.com/science/article/pii/0273117782900709},
author = {J. Trümper}
}

@article{miller2019psr,
  title={PSR J0030+ 0451 mass and radius from NICER data and implications for the properties of neutron star matter},
  author={Miller, MC and Lamb, Frederick K and Dittmann, AJ and Bogdanov, Slavko and Arzoumanian, Zaven and Gendreau, Keith C and Guillot, S and Harding, AK and Ho, WCG and Lattimer, JM and others},
  journal={The Astrophysical Journal Letters},
  volume={887},
  number={1},
  pages={L24},
  year={2019},
  publisher={IOP Publishing}
}

@article{riley2019nicer,
  title={A NICER view of PSR J0030+ 0451: millisecond pulsar parameter estimation},
  author={Riley, Thomas E and Watts, Anna L and Bogdanov, Slavko and Ray, Paul S and Ludlam, Renee M and Guillot, Sebastien and Arzoumanian, Zaven and Baker, Charles L and Bilous, Anna V and Chakrabarty, Deepto and others},
  journal={The Astrophysical Journal Letters},
  volume={887},
  number={1},
  pages={L21},
  year={2019},
  publisher={IOP Publishing}
}

@article{salmi2023atmospheric,
  title={Atmospheric Effects on Neutron Star Parameter Constraints with NICER},
  author={Salmi, Tuomo and Vinciguerra, Serena and Choudhury, Devarshi and Watts, Anna L and Ho, Wynn CG and Guillot, Sebastien and Kini, Yves and Dorsman, Bas and Morsink, Sharon M and Bogdanov, Slavko},
  journal={The Astrophysical Journal},
  volume={956},
  number={2},
  pages={138},
  year={2023},
  publisher={IOP Publishing}
}

@article{vinciguerra2024updated,
  title={An Updated Mass--Radius Analysis of the 2017--2018 NICER Data Set of PSR J0030+ 0451},
  author={Vinciguerra, Serena and Salmi, Tuomo and Watts, Anna L and Choudhury, Devarshi and Riley, Thomas E and Ray, Paul S and Bogdanov, Slavko and Kini, Yves and Guillot, Sebastien and Chakrabarty, Deepto and others},
  journal={The Astrophysical Journal},
  volume={961},
  number={1},
  pages={62},
  year={2024},
  publisher={IOP Publishing}
}

@article{miller2021radius,
  title={The radius of PSR J0740+ 6620 from NICER and XMM-Newton data},
  author={Miller, M Coleman and Lamb, FK and Dittmann, AJ and Bogdanov, S and Arzoumanian, Z and Gendreau, KC and Guillot, S and Ho, WCG and Lattimer, JM and Loewenstein, M and others},
  journal={The Astrophysical Journal Letters},
  volume={918},
  number={2},
  pages={L28},
  year={2021},
  publisher={IOP Publishing}
}

@article{riley2021nicer,
  title={A NICER view of the massive pulsar PSR J0740+ 6620 informed by radio timing and XMM-Newton spectroscopy},
  author={Riley, Thomas E and Watts, Anna L and Ray, Paul S and Bogdanov, Slavko and Guillot, Sebastien and Morsink, Sharon M and Bilous, Anna V and Arzoumanian, Zaven and Choudhury, Devarshi and Deneva, Julia S and others},
  journal={The Astrophysical Journal Letters},
  volume={918},
  number={2},
  pages={L27},
  year={2021},
  publisher={IOP Publishing}
}

@article{salmi2022radius,
  title={The radius of PSR J0740+ 6620 from NICER with NICER background estimates},
  author={Salmi, Tuomo and Vinciguerra, Serena and Choudhury, Devarshi and Riley, Thomas E and Watts, Anna L and Remillard, Ronald A and Ray, Paul S and Bogdanov, Slavko and Guillot, Sebastien and Arzoumanian, Zaven and others},
  journal={The Astrophysical Journal},
  volume={941},
  number={2},
  pages={150},
  year={2022},
  publisher={IOP Publishing}
}

@article{salmi2024radius,
  title={The Radius of the High Mass Pulsar PSR J0740+ 6620 With 3.6 Years of NICER Data},
  author={Salmi, Tuomo and Choudhury, Devarshi and Kini, Yves and Riley, Thomas E and Vinciguerra, Serena and Watts, Anna L and Wolff, Michael T and Arzoumanian, Zaven and Bogdanov, Slavko and Chakrabarty, Deepto and others},
  journal={arXiv preprint arXiv:2406.14466},
  year={2024}
}

@article{dittmann2024more,
  title={A more precise measurement of the radius of psr j0740+ 6620 using updated nicer data},
  author={Dittmann, Alexander J and Miller, M Coleman and Lamb, Frederick K and Holt, Isiah M and Chirenti, Cecilia and Wolff, Michael T and Bogdanov, Slavko and Guillot, Sebastien and Ho, Wynn CG and Morsink, Sharon M and others},
  journal={The Astrophysical Journal},
  volume={974},
  number={2},
  pages={295},
  year={2024},
  publisher={IOP Publishing}
}

@article{choudhury2024nicer,
  title={A nicer view of the nearest and brightest millisecond pulsar: Psr j0437--4715},
  author={Choudhury, Devarshi and Salmi, Tuomo and Vinciguerra, Serena and Riley, Thomas E and Kini, Yves and Watts, Anna L and Dorsman, Bas and Bogdanov, Slavko and Guillot, Sebastien and Ray, Paul S and others},
  journal={The Astrophysical Journal Letters},
  volume={971},
  number={1},
  pages={L20},
  year={2024},
  publisher={IOP Publishing}
}

@article{salmi2024nicer,
  title={A NICER View of PSR J1231- 1411: A Complex Case},
  author={Salmi, Tuomo and Deneva, Julia S and Ray, Paul S and Watts, Anna L and Choudhury, Devarshi and Kini, Yves and Vinciguerra, Serena and Cromartie, H Thankful and Wolff, Michael T and Arzoumanian, Zaven and others},
  journal={The Astrophysical Journal},
  volume={976},
  number={1},
  pages={58},
  year={2024},
  publisher={IOP Publishing}
}

@article{Qi_2025,
doi = {10.3847/1538-4357/adb42f},
url = {https://doi.org/10.3847/1538-4357/adb42f},
year = {2025},
month = {mar},
publisher = {The American Astronomical Society},
volume = {981},
number = {2},
pages = {99},
author = {Qi, Liqiang and Zheng, Shijie and Zhang, Juan and Ge, Mingyu and Li, Ang and Zhang, Shuang-Nan and Lu, Fangjun and Peng, Hanlong and Zhang, Liang and Feng, Hua and Zhang, Zhen and Xu, Yupeng and Li, Zhengwei and Song, Liming and Zhang, Shu and Tao, Lian and Ye, Wentao},
title = {PSR J1231–1411 Revisited: Pulse Profile Analysis of X-Ray Observation},
journal = {The Astrophysical Journal}
}

@article{mauviard2025nicer,
  title={A NICER View of the 1.4 M⊙ Edge-on Pulsar PSR J0614-3329},
  author={Mauviard, Lucien and Guillot, Sebastien and Salmi, Tuomo and Choudhury, Devarshi and Dorsman, Bas and Gonz{\'a}lez-Caniulef, Denis and Hoogkamer, Mariska and Huppenkothen, Daniela and Kazantsev, Christine and Kini, Yves and others},
  journal={The Astrophysical Journal},
  volume={995},
  number={1},
  pages={60},
  year={2025},
  publisher={IOP Publishing}
}

@article{poutanen2003nature,
  title={On the nature of the X-ray emission from the accreting millisecond pulsar SAX J1808. 4- 3658},
  author={Poutanen, Juri and Gierli{\'n}ski, Marek},
  journal={Monthly Notices of the Royal Astronomical Society},
  volume={343},
  number={4},
  pages={1301--1311},
  year={2003},
  publisher={Blackwell Science Ltd Oxford, UK}
}

@article{poutanen2006pulse,
  title={Pulse profiles of millisecond pulsars and their Fourier amplitudes},
  author={Poutanen, Juri and Beloborodov, Andrei M},
  journal={Monthly Notices of the Royal Astronomical Society},
  volume={373},
  number={2},
  pages={836--844},
  year={2006},
  publisher={Blackwell Publishing Ltd Oxford, UK}
}

@article{cadeau2007light,
  title={Light curves for rapidly rotating neutron stars},
  author={Cadeau, Coire and Morsink, Sharon M and Leahy, Denis and Campbell, Sheldon S},
  journal={The Astrophysical Journal},
  volume={654},
  number={1},
  pages={458},
  year={2007},
  publisher={IOP Publishing}
}

@article{morsink2007oblate,
  title={The oblate schwarzschild approximation for light curves of rapidly rotating neutron stars},
  author={Morsink, Sharon M and Leahy, Denis A and Cadeau, Coire and Braga, John},
  journal={The Astrophysical Journal},
  volume={663},
  number={2},
  pages={1244},
  year={2007},
  publisher={IOP Publishing}
}

@article{algendy2014universality,
  title={Universality of the Acceleration due to Gravity on the Surface of a Rapidly Rotating Neutron Star},
  author={AlGendy, Mohammad and Morsink, Sharon M},
  journal={The Astrophysical Journal},
  volume={791},
  number={2},
  pages={78},
  year={2014},
  publisher={IOP Publishing}
}

@article{nattila2018radiation,
  title={Radiation from rapidly rotating oblate neutron stars},
  author={N{\"a}ttil{\"a}, Joonas and Pihajoki, Pauli},
  journal={Astronomy \& Astrophysics},
  volume={615},
  pages={A50},
  year={2018},
  publisher={EDP sciences}
}

@article{bogdanov2019constrainingii,
  title={Constraining the neutron star mass--radius relation and dense matter equation of state with NICER. II. Emission from hot spots on a rapidly rotating neutron star},
  author={Bogdanov, Slavko and Lamb, Frederick K and Mahmoodifar, Simin and Miller, M Coleman and Morsink, Sharon M and Riley, Thomas E and Strohmayer, Tod E and Tung, Albert K and Watts, Anna L and Dittmann, Alexander J and others},
  journal={The Astrophysical Journal Letters},
  volume={887},
  number={1},
  pages={L26},
  year={2019},
  publisher={IOP Publishing}
}

@article{riley2023x,
  title={X-PSI: A Python package for neutron star X-ray pulse simulation and inference},
  author={Riley, Thomas E and Choudhury, Devarshi and Salmi, Tuomo and Vinciguerra, Serena and Kini, Yves and Dorsman, Bas and Watts, Anna L and Huppenkothen, Daniela and Guillot, Sebastien},
  journal={Journal of Open Source Software},
  volume={8},
  year={2023}
}

@article{wilms2000absorption,
  title={On the absorption of X-rays in the interstellar medium},
  author={Wilms, J and Allen, A and McCray, R},
  journal={The Astrophysical Journal},
  volume={542},
  number={2},
  pages={914},
  year={2000},
  publisher={IOP Publishing}
}

@article{feroz2009multinest,
  title={MultiNest: an efficient and robust Bayesian inference tool for cosmology and particle physics},
  author={Feroz, Farhan and Hobson, MP and Bridges, Michael},
  journal={Monthly Notices of the Royal Astronomical Society},
  volume={398},
  number={4},
  pages={1601--1614},
  year={2009},
  publisher={Blackwell Publishing Ltd Oxford, UK}
}

@article{watts2016colloquium,
  title={Colloquium: Measuring the neutron star equation of state using x-ray timing},
  author={Watts, Anna L and Andersson, Nils and Chakrabarty, Deepto and Feroci, Marco and Hebeler, Kai and Israel, Gianluca and Lamb, Frederick K and Miller, M Coleman and Morsink, Sharon and {\"O}zel, Feryal and others},
  journal={Reviews of Modern Physics},
  volume={88},
  number={2},
  pages={021001},
  year={2016},
  publisher={APS}
}

@article{raaijmakers2021constraints,
  title={Constraints on the dense matter equation of state and neutron star properties from NICER’s mass--radius estimate of PSR J0740+ 6620 and multimessenger observations},
  author={Raaijmakers, Geert and Greif, SK and Hebeler, K and Hinderer, T and Nissanke, aS and Schwenk, A and Riley, TE and Watts, AL and Lattimer, JM and Ho, WCG},
  journal={The Astrophysical Journal Letters},
  volume={918},
  number={2},
  pages={L29},
  year={2021},
  publisher={IOP Publishing}
}

@article{miao2024thermal,
  title={Thermal x-ray studies of neutron stars and the equation of state},
  author={Miao, Zhiqiang and Qi, Liqiang and Zhang, Juan and Li, Ang and Ge, Mingyu},
  journal={Physical Review D},
  volume={109},
  number={12},
  pages={123005},
  year={2024},
  publisher={APS}
}

@article{bilous2019nicer,
  title={A NICER view of PSR J0030+ 0451: evidence for a global-scale multipolar magnetic field},
  author={Bilous, Anna V and Watts, Anna L and Harding, Alice K and Riley, Thomas E and Arzoumanian, Zaven and Bogdanov, Slavko and Gendreau, Keith C and Ray, Paul S and Guillot, Sebastien and Ho, Wynn CG and others},
  journal={The Astrophysical Journal Letters},
  volume={887},
  number={1},
  pages={L23},
  year={2019},
  publisher={IOP Publishing}
}

@article{chen2020numerical,
  title={A numerical model for the multiwavelength lightcurves of PSR J0030+ 0451},
  author={Chen, Alexander Y and Yuan, Yajie and Vasilopoulos, Georgios},
  journal={The Astrophysical Journal Letters},
  volume={893},
  number={2},
  pages={L38},
  year={2020},
  publisher={IOP Publishing}
}

@article{kalapotharakos2021multipolar,
  title={The multipolar magnetic field of the millisecond pulsar PSR J0030+ 0451},
  author={Kalapotharakos, Constantinos and Wadiasingh, Zorawar and Harding, Alice K and Kazanas, Demosthenes},
  journal={The Astrophysical Journal},
  volume={907},
  number={2},
  pages={63},
  year={2021},
  publisher={IOP Publishing}
}

@article{carrasco2023relativistic,
  title={Relativistic force-free models of the thermal X-ray emission in millisecond pulsars observed by NICER},
  author={Carrasco, Federico and Pelle, Joaquin and Reula, Oscar and Vigan{\`o}, Daniele and Palenzuela, Carlos},
  journal={Monthly Notices of the Royal Astronomical Society},
  volume={520},
  number={2},
  pages={3151--3163},
  year={2023},
  publisher={Oxford University Press}
}

@article{lattimer2001neutron,
  title={Neutron star structure and the equation of state},
  author={Lattimer, JM and Prakash, M},
  journal={The Astrophysical Journal},
  volume={550},
  number={1},
  pages={426},
  year={2001},
  publisher={IOP Publishing}
}

@misc{J0437data,
  doi = {10.5281/ZENODO.10886504},
  url = {https://zenodo.org/doi/10.5281/zenodo.10886504},
  author = {Choudhury,  Devarshi and Salmi,  Tuomo and Serena,  Vinciguerra and Riley,  Thomas and Kini,  Yves and Watts,  Anna L. and Dorsman,  Bas and Bogdanov,  Slavko and Guillot,  Sebastien and Ray,  Paul S. and Reardon,  Daniel and Remillard,  Ronald A. and Bilous,  Anna and Huppenkothen,  Daniela and Lattimer,  James and Rutherford,  Nathan and Arzoumanian,  Zaven and Gendreau,  Keith and Morsink,  Sharon and Ho,  Wynn C. G.},
  title = {Reproduction package for:  'A NICER View of the Nearest and Brightest Millisecond Pulsar: PSR J0437–4715'},
  publisher = {Zenodo},
  year = {2024},
  copyright = {Creative Commons Attribution 4.0 International}
}

@misc{nsxmodel,
  doi = {10.5281/ZENODO.7094144},
  url = {https://zenodo.org/record/7094144},
  author = {Watts,  Anna L. and Salmi,  Tuomo and Choudhury,  Devarshi and Vinciguerra,  Serena and Kini,  Yves and Dorsman,  Bas and Huppenkothen,  Daniela and Guillot,  Sebastien and Ho,  Wynn},
  language = {en},
  title = {Auxiliary files for X-PSI tutorials},
  publisher = {Zenodo},
  year = {2022},
  copyright = {Creative Commons Attribution 4.0 International}
}

@article{10.1093/mnras/stw2194,
    author = {Guillot, S. and Kaspi, V. M. and Archibald, R. F. and Bachetti, M. and Flynn, C. and Jankowski, F. and Bailes, M. and Boggs, S. and Christensen, F. E. and Craig, W. W. and Hailey, C. A. and Harrison, F. A. and Stern, D. and Zhang, W. W.},
    title = {The NuSTAR view of the non-thermal emission from PSR J0437−4715},
    journal = {Monthly Notices of the Royal Astronomical Society},
    volume = {463},
    number = {3},
    pages = {2612-2622},
    year = {2016},
    month = {09},
    issn = {0035-8711},
    doi = {10.1093/mnras/stw2194},
    url = {https://doi.org/10.1093/mnras/stw2194},
    eprint = {https://academic.oup.com/mnras/article-pdf/463/3/2612/18242835/stw2194.pdf}
}

@article{Durant_2012,
doi = {10.1088/0004-637X/746/1/6},
url = {https://doi.org/10.1088/0004-637X/746/1/6},
year = {2012},
month = {jan},
publisher = {The American Astronomical Society},
volume = {746},
number = {1},
pages = {6},
author = {Durant, Martin and Kargaltsev, Oleg and Pavlov, George G. and Kowalski, Piotr M. and Posselt, Bettina and van Kerkwijk, Marten H. and Kaplan, David L.},
title = {THE SPECTRUM OF THE RECYCLED PSR J0437−4715 AND ITS WHITE DWARF COMPANION},
journal = {The Astrophysical Journal}
}

@article{Abdo_2010,
doi = {10.1088/0067-0049/187/2/460},
url = {https://doi.org/10.1088/0067-0049/187/2/460},
year = {2010},
month = {mar},
publisher = {The American Astronomical Society},
volume = {187},
number = {2},
pages = {460},
author = {Abdo, A. A. and Ackermann, M. and Ajello, M. and Atwood, W. B. and Axelsson, M. and Baldini, L. and Ballet, J. and Barbiellini, G. and Baring, M. G. and Bastieri, D. and Baughman, B. M. and Bechtol, K. and Bellazzini, R. and Berenji, B. and Blandford, R. D. and Bloom, E. D. and Bonamente, E. and Borgland, A. W. and Bregeon, J. and Brez, A. and Brigida, M. and Bruel, P. and Burnett, T. H. and Buson, S. and Caliandro, G. A. and Cameron, R. A. and Camilo, F. and Caraveo, P. A. and Casandjian, J. M. and Cecchi, C. and Çelik, Ö. and Charles, E. and Chekhtman, A. and Cheung, C. C. and Chiang, J. and Ciprini, S. and Claus, R. and Cognard, I. and Cohen-Tanugi, J. and Cominsky, L. R. and Conrad, J. and Corbet, R. and Cutini, S. and den Hartog, P. R. and Dermer, C. D. and de Angelis, A. and de Luca, A. and de Palma, F. and Digel, S. W. and Dormody, M. and do Couto e Silva, E. and Drell, P. S. and Dubois, R. and Dumora, D. and Espinoza, C. and Farnier, C. and Favuzzi, C. and Fegan, S. J. and Ferrara, E. C. and Focke, W. B. and Fortin, P. and Frailis, M. and Freire, P. C. C. and Fukazawa, Y. and Funk, S. and Fusco, P. and Gargano, F. and Gasparrini, D. and Gehrels, N. and Germani, S. and Giavitto, G. and Giebels, B. and Giglietto, N. and Giommi, P. and Giordano, F. and Glanzman, T. and Godfrey, G. and Gotthelf, E. V. and Grenier, I. A. and Grondin, M.-H. and Grove, J. E. and Guillemot, L. and Guiriec, S. and Gwon, C. and Hanabata, Y. and Harding, A. K. and Hayashida, M. and Hays, E. and Hughes, R. E. and Jackson, M. S. and Jóhannesson, G. and Johnson, A. S. and Johnson, R. P. and Johnson, T. J. and Johnson, W. N. and Johnston, S. and Kamae, T. and Kanbach, G. and Kaspi, V. M. and Katagiri, H. and Kataoka, J. and Kawai, N. and Kerr, M. and Knödlseder, J. and Kocian, M. L. and Kramer, M. and Kuss, M. and Lande, J. and Latronico, L. and Lemoine-Goumard, M. and Livingstone, M. and Longo, F. and Loparco, F. and Lott, B. and Lovellette, M. N. and Lubrano, P. and Lyne, A. G. and Madejski, G. M. and Makeev, A. and Manchester, R. N. and Marelli, M. and Mazziotta, M. N. and McConville, W. and McEnery, J. E. and McGlynn, S. and Meurer, C. and Michelson, P. F. and Mineo, T. and Mitthumsiri, W. and Mizuno, T. and Moiseev, A. A. and Monte, C. and Monzani, M. E. and Morselli, A. and Moskalenko, I. V. and Murgia, S. and Nakamori, T. and Nolan, P. L. and Norris, J. P. and Noutsos, A. and Nuss, E. and Ohsugi, T. and Omodei, N. and Orlando, E. and Ormes, J. F. and Ozaki, M. and Paneque, D. and Panetta, J. H. and Parent, D. and Pelassa, V. and Pepe, M. and Pesce-Rollins, M. and Piron, F. and Porter, T. A. and Rainò, S. and Rando, R. and Ransom, S. M. and Ray, P. S. and Razzano, M. and Rea, N. and Reimer, A. and Reimer, O. and Reposeur, T. and Ritz, S. and Rodriguez, A. Y. and Romani, R. W. and Roth, M. and Ryde, F. and Sadrozinski, H. F.-W. and Sanchez, D. and Sander, A. and Saz Parkinson, P. M. and Scargle, J. D. and Schalk, T. L. and Sellerholm, A. and Sgrò, C. and Siskind, E. J. and Smith, D. A. and Smith, P. D. and Spandre, G. and Spinelli, P. and Stappers, B. W. and Starck, J.-L. and Striani, E. and Strickman, M. S. and Strong, A. W. and Suson, D. J. and Tajima, H. and Takahashi, H. and Takahashi, T. and Tanaka, T. and Thayer, J. B. and Thayer, J. G. and Theureau, G. and Thompson, D. J. and Thorsett, S. E. and Tibaldo, L. and Tibolla, O. and Torres, D. F. and Tosti, G. and Tramacere, A. and Uchiyama, Y. and Usher, T. L. and Van Etten, A. and Vasileiou, V. and Venter, C. and Vilchez, N. and Vitale, V. and Waite, A. P. and Wang, P. and Wang, N. and Watters, K. and Weltevrede, P. and Winer, B. L. and Wood, K. S. and Ylinen, T. and Ziegler, M.},
title = {THE FIRST FERMI LARGE AREA TELESCOPE CATALOG OF GAMMA-RAY PULSARS},
journal = {The Astrophysical Journal Supplement Series}
}

@article{Bogdanov_2013,
doi = {10.1088/0004-637X/762/2/96},
url = {https://doi.org/10.1088/0004-637X/762/2/96},
year = {2012},
month = {dec},
publisher = {The American Astronomical Society},
volume = {762},
number = {2},
pages = {96},
author = {Bogdanov, Slavko},
title = {THE NEAREST MILLISECOND PULSAR REVISITED WITH XMM-NEWTON: IMPROVED MASS–RADIUS CONSTRAINTS FOR PSR J0437–4715},
journal = {The Astrophysical Journal}
}

@article{10.1093/mnras/stz2941,
    author = {González-Caniulef, Denis and Guillot, Sebastien and Reisenegger, Andreas},
    title = {Neutron star radius measurement from the ultraviolet and soft X-ray thermal emission of PSR J0437−4715},
    journal = {Monthly Notices of the Royal Astronomical Society},
    volume = {490},
    number = {4},
    pages = {5848-5859},
    year = {2019},
    month = {11},
    issn = {0035-8711},
    doi = {10.1093/mnras/stz2941},
    url = {https://doi.org/10.1093/mnras/stz2941},
    eprint = {https://academic.oup.com/mnras/article-pdf/490/4/5848/31084725/stz2941.pdf},
}

@article{Reardon_2024,
doi = {10.3847/2041-8213/ad614a},
url = {https://doi.org/10.3847/2041-8213/ad614a},
year = {2024},
month = {aug},
publisher = {The American Astronomical Society},
volume = {971},
number = {1},
pages = {L18},
author = {Reardon, Daniel J. and Bailes, Matthew and Shannon, Ryan M. and Flynn, Chris and Askew, Jacob and Bhat, N. D. Ramesh and Chen, Zu-Cheng and Curyło, Małgorzata and Feng, Yi and Hobbs, George B. and Kapur, Agastya and Kerr, Matthew and Liu, Xiaojin and Manchester, Richard N. and Mandow, Rami and Mishra, Saurav and Russell, Christopher J. and Shamohammadi, Mohsen and Zhang, Lei and Zic, Andrew},
title = {The Neutron Star Mass, Distance, and Inclination from Precision Timing of the Brilliant Millisecond Pulsar J0437-4715},
journal = {The Astrophysical Journal Letters}
}

@ARTICLE{1995ApJ...441L..65M,
       author = {{Manchester}, R.~N. and {Johnston}, Simon},
        title = "{Polarization Properties of Two Pulsars}",
      journal = {\apjl},
         year = 1995,
        month = mar,
       volume = {441},
        pages = {L65},
          doi = {10.1086/187791},
       adsurl = {https://ui.adsabs.harvard.edu/abs/1995ApJ...441L..65M}
}

@article{Bhat_2014,
doi = {10.1088/2041-8205/791/2/L32},
url = {https://doi.org/10.1088/2041-8205/791/2/L32},
year = {2014},
month = {aug},
publisher = {The American Astronomical Society},
volume = {791},
number = {2},
pages = {L32},
author = {Bhat, N. D. R. and Ord, S. M. and Tremblay, S. E. and Tingay, S. J. and Deshpande, A. A. and van Straten, W. and Oronsaye, S. and Bernardi, G. and Bowman, J. D. and Briggs, F. and Cappallo, R. J. and Corey, B. E. and Emrich, D. and Goeke, R. and Greenhill, L. J. and Hazelton, B. J. and Hewitt, J. N. and Johnston-Hollitt, M. and Kaplan, D. L. and Kasper, J. C. and Kratzenberg, E. and Lonsdale, C. J. and Lynch, M. J. and McWhirter, S. R. and Mitchell, D. A. and Morales, M. F. and Morgan, E. and Oberoi, D. and Prabu, T. and Rogers, A. E. E. and Roshi, D. A. and Udaya Shankar, N. and Srivani, K. S. and Subrahmanyan, R. and Waterson, M. and Wayth, R. B. and Webster, R. L. and Whitney, A. R. and Williams, A. and Williams, C. L.},
title = {THE LOW-FREQUENCY CHARACTERISTICS OF PSR J0437−4715 OBSERVED WITH THE MURCHISON WIDE-FIELD ARRAY},
journal = {The Astrophysical Journal Letters}
}

@article{OLSON1987325,
title = {Short characteristic solution of the non-LTE line transfer problem by operator perturbation—I. The one-dimensional planar slab},
journal = {Journal of Quantitative Spectroscopy and Radiative Transfer},
volume = {38},
number = {5},
pages = {325-336},
year = {1987},
issn = {0022-4073},
doi = {https://doi.org/10.1016/0022-4073(87)90027-6},
url = {https://www.sciencedirect.com/science/article/pii/0022407387900276},
author = {Gordon L. Olson and P.B. Kunasz}
}

@BOOK{1978stat.book.....M,
author = {{Mihalas}, Dimitri},
title = "{Stellar atmospheres}",
year = 1978,
adsurl = {https://ui.adsabs.harvard.edu/abs/1978stat.book.....M}
}

@article{Clayton_2003,
doi = {10.1086/374316},
url = {https://doi.org/10.1086/374316},
year = {2003},
month = {may},
publisher = {},
volume = {588},
number = {2},
pages = {871},
author = {Clayton, Geoffrey C. and Wolff, Michael J. and Sofia, Ulysses J. and Gordon, K. D. and Misselt, K. A.},
title = {Dust Grain Size Distributions from MRN to MEM},
journal = {The Astrophysical Journal}
}

@ARTICLE{1990ApJS...72..163F,
       author = {{Fitzpatrick}, Edward L. and {Massa}, Derck},
        title = "{An Analysis of the Shapes of Ultraviolet Extinction Curves. III. an Atlas of Ultraviolet Extinction Curves}",
      journal = {\apjs},
         year = 1990,
        month = jan,
       volume = {72},
        pages = {163},
          doi = {10.1086/191413},
       adsurl = {https://ui.adsabs.harvard.edu/abs/1990ApJS...72..163F}
}

@ARTICLE{2018AA...616A.132L,
       author = {{Lallement}, R. and {Capitanio}, L. and {Ruiz-Dern}, L. and {Danielski}, C. and {Babusiaux}, C. and {Vergely}, L. and {Elyajouri}, M. and {Arenou}, F. and {Leclerc}, N.},
        title = "{Three-dimensional maps of interstellar dust in the Local Arm: using Gaia, 2MASS, and APOGEE-DR14}",
      journal = {\aap},
         year = 2018,
        month = aug,
       volume = {616},
          eid = {A132},
        pages = {A132},
          doi = {10.1051/0004-6361/201832832},
archivePrefix = {arXiv},
       eprint = {1804.06060},
 primaryClass = {astro-ph.GA},
       adsurl = {https://ui.adsabs.harvard.edu/abs/2018A&A...616A.132L}
}

@article{10.1093/mnras/stz2524,
    author = {Lockhart, Will and Gralla, Samuel E and Özel, Feryal and Psaltis, Dimitrios},
    title = {X-ray light curves from realistic polar cap models: inclined pulsar magnetospheres and multipole fields},
    journal = {Monthly Notices of the Royal Astronomical Society},
    volume = {490},
    number = {2},
    pages = {1774-1783},
    year = {2019},
    month = {09},
    issn = {0035-8711},
    doi = {10.1093/mnras/stz2524},
    url = {https://doi.org/10.1093/mnras/stz2524},
    eprint = {https://academic.oup.com/mnras/article-pdf/490/2/1774/30194725/stz2524.pdf},
}

@ARTICLE{2024Univ...10..174Z,
       author = {{Zheng}, Shijie and {Han}, Dawei and {Xu}, Heng and {Lee}, Kejia and {Yuan}, Jianping and {Wang}, Haoxi and {Ge}, Mingyu and {Zhang}, Liang and {Li}, Yongye and {Yin}, Yitao and {Ma}, Xiang and {Chen}, Yong and {Zhang}, Shuangnan},
        title = "{New Timing Results of MSPs from NICER Observations}",
      journal = {Universe},
         year = 2024,
        month = apr,
       volume = {10},
       number = {4},
          eid = {174},
        pages = {174},
          doi = {10.3390/universe10040174},
archivePrefix = {arXiv},
       eprint = {2404.16263},
 primaryClass = {astro-ph.HE},
       adsurl = {https://ui.adsabs.harvard.edu/abs/2024Univ...10..174Z}
}

@ARTICLE{2019MNRAS.490.4666P,
       author = {{Perera}, B.~B.~P. and {DeCesar}, M.~E. and {Demorest}, P.~B. and {Kerr}, M. and {Lentati}, L. and {Nice}, D.~J. and {Os{\l}owski}, S. and {Ransom}, S.~M. and {Keith}, M.~J. and {Arzoumanian}, Z. and {Bailes}, M. and {Baker}, P.~T. and {Bassa}, C.~G. and {Bhat}, N.~D.~R. and {Brazier}, A. and {Burgay}, M. and {Burke-Spolaor}, S. and {Caballero}, R.~N. and {Champion}, D.~J. and {Chatterjee}, S. and {Chen}, S. and {Cognard}, I. and {Cordes}, J.~M. and {Crowter}, K. and {Dai}, S. and {Desvignes}, G. and {Dolch}, T. and {Ferdman}, R.~D. and {Ferrara}, E.~C. and {Fonseca}, E. and {Goldstein}, J.~M. and {Graikou}, E. and {Guillemot}, L. and {Hazboun}, J.~S. and {Hobbs}, G. and {Hu}, H. and {Islo}, K. and {Janssen}, G.~H. and {Karuppusamy}, R. and {Kramer}, M. and {Lam}, M.~T. and {Lee}, K.~J. and {Liu}, K. and {Luo}, J. and {Lyne}, A.~G. and {Manchester}, R.~N. and {McKee}, J.~W. and {McLaughlin}, M.~A. and {Mingarelli}, C.~M.~F. and {Parthasarathy}, A.~P. and {Pennucci}, T.~T. and {Perrodin}, D. and {Possenti}, A. and {Reardon}, D.~J. and {Russell}, C.~J. and {Sanidas}, S.~A. and {Sesana}, A. and {Shaifullah}, G. and {Shannon}, R.~M. and {Siemens}, X. and {Simon}, J. and {Spiewak}, R. and {Stairs}, I.~H. and {Stappers}, B.~W. and {Swiggum}, J.~K. and {Taylor}, S.~R. and {Theureau}, G. and {Tiburzi}, C. and {Vallisneri}, M. and {Vecchio}, A. and {Wang}, J.~B. and {Zhang}, S.~B. and {Zhang}, L. and {Zhu}, W.~W. and {Zhu}, X.~J.},
        title = "{The International Pulsar Timing Array: second data release}",
      journal = {\mnras},
         year = 2019,
        month = dec,
       volume = {490},
       number = {4},
        pages = {4666-4687},
          doi = {10.1093/mnras/stz2857},
archivePrefix = {arXiv},
       eprint = {1909.04534},
 primaryClass = {astro-ph.HE},
       adsurl = {https://ui.adsabs.harvard.edu/abs/2019MNRAS.490.4666P}
}

@software{1999ascl.soft10005A,
       author = {{Arnaud}, Keith and {Dorman}, Ben and {Gordon}, Craig},
        title = "{XSPEC: An X-ray spectral fitting package}",
 howpublished = {Astrophysics Source Code Library, record ascl:9910.005},
         year = 1999,
        month = oct,
          eid = {ascl:9910.005},
archivePrefix = {ascl},
       eprint = {9910.005},
       adsurl = {https://ui.adsabs.harvard.edu/abs/1999ascl.soft10005A}
}

@ARTICLE{gaia,
       author = {{Vergely}, J.~L. and {Lallement}, R. and {Cox}, N.~L.~J.},
        title = "{Three-dimensional extinction maps: Inverting inter-calibrated extinction catalogues}",
      journal = {\aap},
         year = 2022,
        month = aug,
       volume = {664},
          eid = {A174},
        pages = {A174},
          doi = {10.1051/0004-6361/202243319},
archivePrefix = {arXiv},
       eprint = {2205.09087},
 primaryClass = {astro-ph.GA},
       adsurl = {https://ui.adsabs.harvard.edu/abs/2022A&A...664A.174V}
}

@unpublished{Zheng2026prep,
  author  = {Zheng, S.~J.},
  title   = {in preparation},
  year = {\ndd},
  note    = {in preparation},
}

@article{10.1093/mnras/stae1175,
    author = {Johnston, Simon and Mitra, Dipanjan and Keith, Michael J and Oswald, Lucy S and Karastergiou, Aris},
    title = {The Thousand-Pulsar-Array programme on MeerKAT – XIV. On the high linearly polarized pulsar signals},
    journal = {Monthly Notices of the Royal Astronomical Society},
    volume = {530},
    number = {4},
    pages = {4839-4849},
    year = {2024},
    month = {04},
    issn = {0035-8711},
    doi = {10.1093/mnras/stae1175},
    url = {https://doi.org/10.1093/mnras/stae1175},
    eprint = {https://academic.oup.com/mnras/article-pdf/530/4/4839/57503907/stae1175.pdf},
}

@article{petri2026multi,
  title={Multi-wavelength emission modelling of PSR\~{} J0437 $-$4715},
  author={P{\'e}tri, J and Stammler, P and Guillemot, L and Guillot, S and Gonz{\'a}lez-Caniulef, D and Jankowski, F and Webb, N},
  journal={arXiv preprint arXiv:2603.10536},
  year={2026}
}

@article{pavlov1997mass,
  title={Mass-to-radius ratio for the millisecond pulsar J0437--4715},
  author={Pavlov, GG and Zavlin, VE},
  journal={The Astrophysical Journal Letters},
  volume={490},
  number={1},
  pages={L91--L94},
  year={1997}
}

@article{Zavlin1998,
  title={Soft X-rays from polar caps of the millisecond pulsar J0437--4715},
  author={Zavlin, VE and Pavlov, GG},
  journal={Astron. Astrophy},
  volume={329},
  pages={583-598},
  year={1998}
}
\bibliographystyle{aasjournal}

\end{document}